\documentclass[a4paper,11pt]{article}
\pdfoutput=1

\newcommand{\BsMassToleLow}{5.28} 
\newcommand{\BsMassToleHigh}{5.46} 
\newcommand{\DsMassToleLow}{1.96} 
\newcommand{\DsMassToleHigh}{1.98} 
\newcommand{\totBsEvents}{2\times10^5} 
\newcommand{\DsPDGMass}{1.9685}
\newcommand{\BsPDGMass}{5.3693}

\newcommand{\gammaAnaErr}{3.01}
\newcommand{\gammaFinalErr}{0.69} 

\usepackage{jheppub}
\usepackage[T1]{fontenc}

\usepackage{HEP} 
\usepackage{braket}
\usepackage{lmodern}

\graphicspath{{figures/}}

\title{\boldmath Prospect for measurement of $C\!P$-violating observables in $B_s^0 \to D_s^\mp K^\pm$ decays at a future ${Z}$ factory}

\author[a,b]{Ji Peng$^*$}
\author[c]{Mingrui Zhao}
\author[d]{Xiaolin Wang}
\author[a]{Manqi Ruan}
\author[d]{Hengne Li}
\author[a]{Shanzhen Chen$^*$}

\affiliation[a]{Institute of High Energy Physics, Chinese Academy of Sciences, Beijing 100049, China}
\affiliation[b]{University of Chinese Academy of Sciences (UCAS), Beijing 100049, China}
\affiliation[c]{Science and Technology on Nuclear Data Laboratory, China Institute of Atomic Energy, Beijing 102413, China}
\affiliation[d]{South China Normal University (SCNU), Guangzhou 510631, China}

\emailAdd{pengji@ihep.ac.cn}
\emailAdd{mingrui.zhao@cern.ch}
\emailAdd{xiaolin.wang@cern.ch}
\emailAdd{ruanmq@ihep.ac.cn}
\emailAdd{hengne.li@cern.ch}

\emailAdd{shanzhen.chen@ihep.ac.cn}

\note[*]{Corresponding authors.}

\abstract{
A precise determination of the CKM angle $\gamma$ from $B_s^0$ oscillations in $B_s^0 \to D_s^\mp K^\pm$ decays offers a critical test of the Standard Model and probes for new physics. We present a comprehensive study on the prospects of measuring $\gamma$ at a future Tera-$Z$ factory, utilizing the baseline detector concept of the Circular Electron Positron Collider (CEPC). A two-dimensional simultaneous fit framework, incorporating flavor tagging, decay time resolution modeling, and acceptance corrections, is developed using full Monte Carlo simulations of $B_s^0 \to D_s^\mp \left(\to K^\mp K^\pm \pi^\mp\right) K^\pm$ decays and inclusive background processes. The effective flavor tagging power reaches 23.6\%, while the decay time resolution is determined to be $26\mathrm{\,fs}$. Projecting to full statistics of signal events across three dominant $D_s^-$ decay channels, we estimate a statistical precision of $\sigma(\gamma) = \gammaFinalErr^\circ$, which corresponds to 4.1 Tera-$Z$ boson equivalent data. This study has established the feasibility of sub-degree level $\gamma$ measurements at a $Z$ factory, highlighting its unique advantages in time-dependent $C\!P$ violation studies through ultra-precise vertexing and background suppression capabilities.

}

\begin{document} 

\maketitle
\flushbottom

\section{Introduction}
\label{sec:Introduction}

\begin{figure}[h]
\centering
\includegraphics[width=1\textwidth]{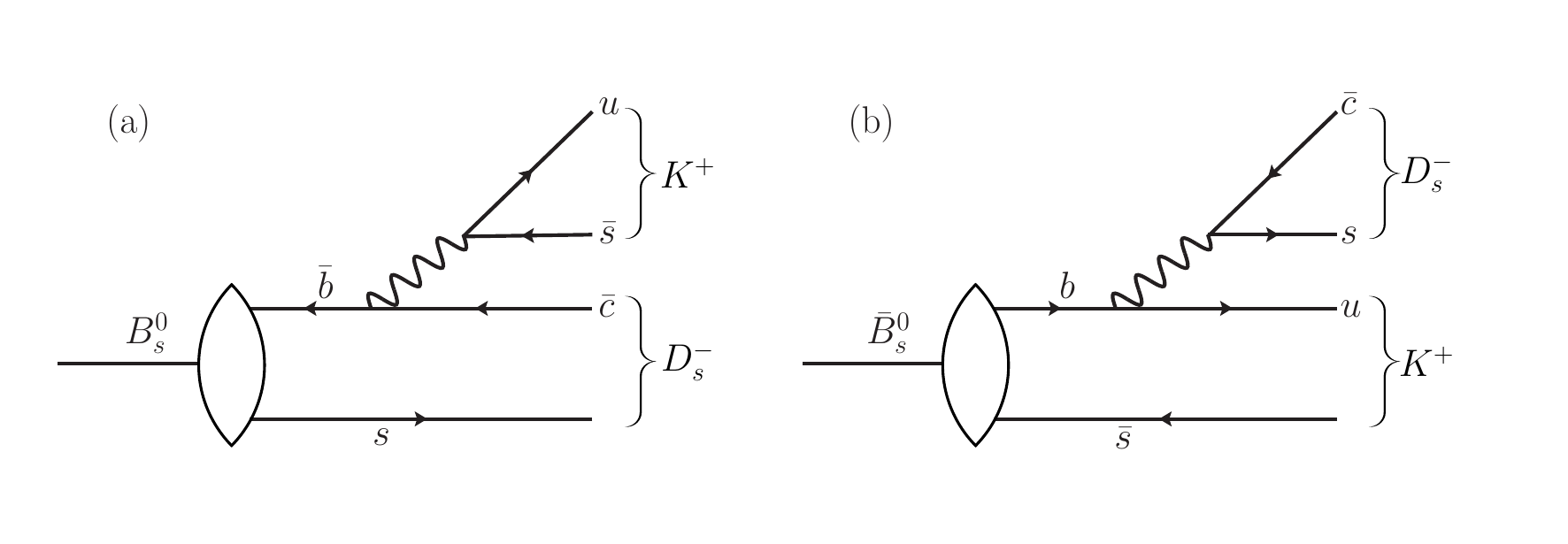}
\caption{Feynman diagrams for (a) $B_s^0\rightarrow D_s^- K^+$ and (b) $\overline{B}_s^0\rightarrow D_s^- K^+$ decays. Time-dependent fits to these two decay channels and their charge conjugate decays allow to determine the angle $\gamma$.}
\label{fig:Diagrams}
\end{figure}

 Charge-parity ($C\!P$) violation is one of the most profound phenomena in particle physics. Searching for the source of $C\!P$ violation is essential for understanding the matter-antimatter asymmetry of the universe~\cite{Bigi:2000yz,Dine:2003ax}. 
 Within the Standard Model (SM), $C\!P$ violation originates from a single complex phase in the Cabibbo-Kobayashi-Maskawa (CKM) matrix~\cite{Cabibbo:1963yz,Kobayashi:1973fv}. The unitarity constraints of this matrix can be represented as triangles in a complex plane, which are known as the Unitarity Triangles. 
The most commonly used unitarity triangle arises from 
$ V_{ud}V^{*}_{ub} + V_{cd}V^{*}_{cb} + V_{td}V^{*}_{tb} = 0$~\cite{Buras:2002yj},
where the internal angle \(\gamma \equiv \arg \left[-V_{ud} V^*_{ub} / V_{cd} V^*_{cb}\right] \) is the least well measured angle, and can be probed directly from tree-level decay processes, or indirectly under the assumption of unitarity. The consistency between direct and indirect measurements of $\gamma$ serves as a test of the SM and a probe for New Physics (NP)~\cite{Fleischer:2021cct, Fleischer:2021cwb}. 
The direct measurement of  \( \gamma \) can be achieved by studying the interference between $b \rightarrow u\overline{c}s$ and $b \rightarrow c\overline{u}s$ transitions, like $B^- \rightarrow \overline{D}^{0} K^{-}$ and $B^- \rightarrow D^{0} K^{-}$ decays\footnote{
Inclusion of charge-conjugate modes is implied throughout except where explicitly stated.} with $D^{0}$ and $\overline{D}^{0}$ decays to common final states.  The precision of such a measurement depends sensitively on the ratio of the interfering amplitudes, which can be generically defined as $r \equiv \left| {A(b \rightarrow u\overline{c}s)}/{A(b \rightarrow c\overline{u}s)} \right|$. In the case of charged \( B \) decays, the relevant ratio is $r_B \equiv |A(B^- \rightarrow \overline{D}^0 K^-) / A(B^- \rightarrow D^0 K^-) |$, which is approximately 0.1~\cite{LHCb:2020yot}. 
Another method to determine \( \gamma \) is through time-dependent $B_s^0 \rightarrow D_s^{\mp}K^{\pm}$ decays  (Fig.~\ref{fig:Diagrams})~\cite{Aleksan1992DeterminingTP,Fleischer:2003yb,DeBruyn:2012jp}. These channels have interferences through $B_s^0 - \overline{B}_s^0$ mixing, therefore it is possible to extract $\gamma - 2\beta_s$, where $\beta_s = \arg \left[-V_{ts} V^*_{tb} / V_{cs} V^*_{cb}\right] $ is related to the phase of $B_s^0$ mixing~\cite{Satta:2008das}. 
The amplitude ratio in  $B_s^0 \rightarrow D_s^{\mp}K^{\pm}$ decays, $r_{D_s K} \equiv \left|A(\overline{B}_s^0 \rightarrow D_s^- K^+)/A(B_s^0 \rightarrow D_s^- K^+)\right|\simeq 0.3$~\cite{LHCb:2024xyw}, is much larger than $r_B$. With sufficient statistics, it is possible to access promising precision.

The Circular Electron Positron Collider (CEPC)~\cite{CEPCStudyGroup:2018ghi} is a proposed next-generation electron-positron collider. It can operate near the $Z$ boson mass energy, and produces trillions of $Z$ bosons, functioning as a Tera-$Z$ factory. Given that the branching ratios of $Z$ boson decays to charm or bottom quark pairs are ${\cal B}$($Z\to c\overline{c})\simeq 12\%$ and ${\cal B}$($Z\to b\overline{b})\simeq 15\%$~\cite{PDG2024} in the SM, respectively, CEPC could also serve as a heavy flavor factory.
In the operation scenario of $50\mathrm{\,MW}$ synchrotron radiation (SR) power per beam~\cite{CEPCStudyGroup:2023quu}, CEPC is expected to produce $4.1\times 10^{12}$ $Z$ bosons within 2 years of $Z$-pole operation mode. In this case, it is expected that approximately $1.2 \times 10^{11}$ $B_s^0$ and $\overline{B}_s^0$ mesons will be produced at CEPC~\cite{Ai:2024nmn}. 

One notable feature of CEPC accelerator is that it can switch the collision energies among the Higgs ($e^+e^- \rightarrow ZH$, $240\mathrm{\,GeV}$), $Z$ ($e^+e^- \rightarrow Z$, $91\mathrm{\,GeV}$), and $WW$ ($e^+e^- \rightarrow W^+W^-$, $160\mathrm{\,GeV}$) modes without requiring hardware changes~\cite{CEPCStudyGroup:2023quu}. This means that the design of CEPC detector has to consider physics topics that can be studied at different collision energies, including Higgs physics, electroweak physics, flavor physics, {\it etc}. In order to study flavor physics, the detector must have excellent particle identification capability for relatively low-momentum particles, precise particle momentum and energy measurements, and precise decay vertex resolution to determine the lifetimes of heavy flavor particles. $B_s^0$ particles can spontaneously oscillate to their antiparticles through the exchange of two $W$ bosons, and their oscillation period are much shorter than their lifetime. Thus, to reconstruct $B_s^0$ decays, it is crucial to measure the lifetime of $B_s^0$ particles, identify their initial flavors, and precise measurements of energies and momenta of final state particles.
Considering the importance of  \( \gamma \) measurements in flavor physics, and the relatively large ratio of the interfering amplitudes, the study of  \( \gamma \) precision with $B_s^0 \rightarrow D_s^{\mp}K^{\pm}$ decays can serve as a benchmark for flavor physics performance at the CEPC detector. 

The  $B_s^0 \rightarrow D_s^{\mp}K^{\pm}$  decay was firstly observed by the CDF collaboration~\cite{CDF:2008gud}, and the $C\!P$ asymmetries in such decays were measured by the LHCb collaboration~\cite{LHCb:2024xyw,LHCb:2017hkl}.
Currently, the most precise determination of $\gamma$ in $B_s^0$ meson decays is reported by LHCb using Run 1 and Run 2 data, giving $\gamma = (81^{+12}_{-11})^\circ$ from $B_s^0 \to D_s^\mp K^\pm$ decays~\cite{LHCb:2024xyw}. However, this precision is limited by complicated backgrounds in hadronic environments, which reduce signal purity and hinder efficient signal selection. In contrast, a Tera-$Z$ factory provides a clean $e^+e^-$ collision environment, leading to enhanced signal purity due to a reduced hadronic background. Additionally, the improved tagging power of $23.6\%$ (see Sec.~\ref{sec:tagging}) at the CEPC, compared to $\sim6\%$ at the LHCb~\cite{Fuhring:2920251}, allows for better identification of the $B_s^0$ meson flavor, significantly improving measurement precision and surpassing the current LHCb results.

To place our study in a broader context, we compare our expected sensitivity with those of current and future experiments. 
The world average from tree-level decays,  as summarized by the HFLAV group in early 2024, gives $\gamma = (66.4^{+2.8}_{-3.0})^\circ$~\cite{HFLAV2024}. More recently, the LHCb collaboration, by combining all available measurements including several new results published in 2023 and 2024, obtains $\gamma = (64.6 \pm 2.8)^\circ$~\cite{LHCb:2024yxi}, representing the most accurate determination to date by a single experiment. Looking ahead, LHCb Upgrade I is expected to collect an integrated luminosity of 50 fb$^{-1}$, which will reduce the uncertainty on $\gamma$ to about $1^\circ$. The subsequent Upgrade II, targeting 300 fb$^{-1}$, aims to reach a precision of approximately $0.35^\circ$~\cite{LHCb:2018roe}. Specifically, for the $B_s^0 \rightarrow D_s^\mp K^\pm$ decay mode, the anticipated precision is $2.5^\circ$ after Upgrade I and $1^\circ$ after Upgrade II. Meanwhile, the Belle II experiment aims to accumulate 50 ab$^{-1}$ of data, with a projected $\gamma$ precision around $1.5^\circ$~\cite{URQUIJO201515}. These anticipated precisions provide a valuable comparative basis for assessing the physics potential of future high-luminosity experiments.

This article presents a study on the prospects of measuring  \( \gamma \) using $B_s^0 \rightarrow D_s^{\mp}K^{\pm}$ decays at a Tera-\( Z \) factory, incorporating full detector simulations and realistic background modeling. We develop a two-dimensional fit framework that simultaneously extracts $C\!P$ observables from mass and decay time distributions, accounting for flavor tagging efficiencies, time-dependent acceptance effects, and resolution smearing. Our results demonstrate that with the sum of signal events from three dominant \( D_{s}^- \) decay channels, a statistical precision of \( \sigma(\gamma) = \gammaFinalErr^\circ \) can be achieved. 
This precision, combined with the prospect of measuring \( \gamma \) in other channels at the CEPC, would rival the most ambitious projections from LHCb and Belle II, establishing it as a next-generation laboratory for precision flavor physics from tree-level processes.

\section{Decay Rate Formalism}
\label{sec:rate_equations}
The time-dependent behavior of the decay \( B_s^0 \to D_s^\mp K^\pm \) is vital for extracting the angle \( \gamma \), necessitating the establishment of a formalism for the decay rate equations.

Following the formalism in Ref.~\cite{Bigi:2000yz}, the time evolution of the $B_s^0$ and $\overline{B}_s^0$ states is governed by
\begin{eqnarray}
    \ket{B_{s}^0(t)} &=&  g_{+}(t)\ket{B_s^0} + \frac{q}{p}g_{-}(t)\ket{\overline{B}_s^0},\notag\\
    \ket{\overline{B}_{s}^0(t)} &=&  g_{+}(t)\ket{\overline{B}_s^0} + \frac{p}{q}g_{-}(t)\ket{{B}_s^0},
\end{eqnarray}
where the time-dependent coefficients are defined as
\begin{eqnarray}
    g_{\pm}(t) &=& \frac{1}{2}e^{-iM_st}e^{-\frac{1}{2}\Gamma_s t}\left[e^{i\frac{\Delta m_s}{2}t}e^{-\frac{\Delta \Gamma_s}{4}t}\pm e^{-i\frac{\Delta m_s}{2}t}e^{\frac{\Delta \Gamma_s}{4}t}\right],
\end{eqnarray}
and the physical parameters are expressed in terms of the mass eigenstates ($B^0_{s,L}$, $B^0_{s,H}$):
\begin{eqnarray} M_s &\equiv& \frac{M_{s,L} + M_{s,H}}{2}, \quad \Gamma_s \equiv \frac{\Gamma_{s,L} + \Gamma_{s,H}}{2}, \notag\\ \Delta m_s &\equiv& M_{s,H} - M_{s,L}, \quad \Delta \Gamma_s \equiv \Gamma_{s,L} - \Gamma_{s,H}, \end{eqnarray}
with $M_{s,L}$, $M_{s,H}$, $\Gamma_{s,L}$, and $\Gamma_{s,H}$ denoting the masses and decay widths of the light (L) and heavy (H) eigenstates, respectively.

Within the Standard Model (SM), the $\frac{q}{p}$ is approximated to excellent accuracy at sub per mille level~\cite{Bigi:2000yz} as
\begin{eqnarray}
     \left(\frac{q}{p}\right)_{B_s^0} = \frac{V^{*}_{tb}V_{ts}}{V_{tb}V^{*}_{ts}} = e^{2i\beta_s}.
\end{eqnarray}

The decay amplitudes for the $B_s^0 \to D_s^\mp K^\pm$ processes, determined from the Feynman diagrams in Fig.~\ref{fig:Diagrams}, are parameterized as
\begin{eqnarray}
     \bra{D_s^- K^+}\mathcal{H}\ket{B_s^0}  &=&  e^{i\delta_{-}}V^{*}_{cb}V_{us}A_{-} \equiv A,\notag\\
     \bra{D_s^+ K^-}\mathcal{H}\ket{\overline{B}_s^0}  &=& e^{i\delta_{-}}V_{cb}V^{*}_{us}A_{-} = A,\notag\\
     \bra{D_s^+ K^-}\mathcal{H}\ket{B_s^0}  &=& e^{i\delta_{+}}V^{*}_{ub}V_{cs}A_{+} = A~ r_{D_s K}~ e^{i(\delta + \gamma)}, \notag\\
     \bra{D_s^- K^+}\mathcal{H}\ket{\overline{B}_s^0}  &=&e^{i\delta_{+}}V_{ub}V^{*}_{cs}A_{+} =  A~ r_{D_s K}~ e^{i(\delta - \gamma)}, 
\end{eqnarray}
where $\delta \equiv \delta_{+} - \delta_{-}$ represents the strong phase difference.

As an illustrative case, consider the time-dependent decay rate for $B_s^0 \to D_s^+ K^-$:
\begin{eqnarray}
    P_{B_s^0 \rightarrow D_s^+ K^-}(t) &=& \left|\bra{D_s^+K^-}\mathcal{H}\ket{B_s^0(t)}\right|^2 \notag \\
    & = & \left|g_{+}(t)\bra{D_s^+K^-}\mathcal{H}\ket{B_s^0} + \frac{q}{p}g_{-}(t)\bra{D_s^+K^-}\mathcal{H}\ket{\overline{B}_s^0}\right|^2 \notag \\
    &=& \left|g_{+}(t)A~ r_{D_s K}~ e^{i(\delta + \gamma)} + e^{2i\beta_s}g_{-}(t)A\right|^2. 
\end{eqnarray}

After explicit evaluation of the modulus squared and reparameterization, the time-dependent probability distribution functions for all four $B_s^0 \to D_s^\pm K^\mp$ decay channels can be expressed in a compact form:
\begin{equation}
\begin{pmatrix} P_{B_s^0 \rightarrow D_s^+ K^-}(t)\\ P_{\overline{B}_s^0 \rightarrow D_s^+ K^-}(t)  \\P_{B_s^0 \rightarrow D_s^- K^+}(t) \\  P_{\overline{B}_s^0 \rightarrow D_s^- K^+}(t)\end{pmatrix} \propto e^{-\Gamma_s \, t} \begin{pmatrix} 1 & - C & D_{\overline{f}} & - S_{\overline{f}} \\  1 & C & D_{\overline{f}} & S_{\overline{f}} \\1 & C & D_f & -S_f \\ 1 & -C & D_f & S_f \end{pmatrix}\begin{pmatrix} \cosh \left( \frac{\Delta \Gamma_s}{2} \, t \right)\\ \cos \left( \Delta m_s \, t \right) \\ \sinh \left( \frac{\Delta \Gamma_s}{2} \, t \right) \\  \sin \left( \Delta m_s \, t \right)\end{pmatrix}.
\label{eq:ideal_decay}
\end{equation}

These decay time distributions are parametrized by 5 $C\!P$ observables $C$, $D_f$, $D_{\overline{f}}$, $S_f$ and $S_{\overline{f}}$, which then are related to the magnitude of the amplitude ratio $r_{D_s K}$, the strong phase difference $\delta$, and the weak phase difference $\gamma -2\beta_s$ by the following equations:
\begin{eqnarray}
    C = \frac{1 - r_{D_s K}^2}{1 + r_{D_s K}^2 },
\end{eqnarray}
\begin{eqnarray}
    D_f = \frac{-2\,r_{D_s K}\,\cos \left(\delta -(\gamma-2\,\beta_s)   \right)}{1 + r_{D_s K}^2 }, \qquad
    D_{\overline{f}} = \frac{-2\,r_{D_s K}\,\cos \left(\delta +(\gamma-2\,\beta_s) \right)}{1 + r_{D_s K}^2 },
\end{eqnarray}
\begin{eqnarray}
    S_f = \frac{2\,r_{D_s K}\,\sin \left(\delta -(\gamma-2\,\beta_s) \right)}{1 + r_{D_s K}^2 } ,\qquad
    S_{\overline{f}} = \frac{-2\,r_{D_s K}\,\sin \left(\delta +(\gamma-2\,\beta_s) \right)}{1 + r_{D_s K}^2 }.
\end{eqnarray}

With fits to the decay time distributions of the samples obtained from CEPC, a measurement of \(\gamma\) can be performed using an independent measurement of \(-2\beta_s\) as input.

\section{Experimental Framework}
\subsection{Detector and simulation}
\label{sec:Chan}

\begin{figure}[htb]
\centering
\centering\includegraphics[width=0.45\textwidth]{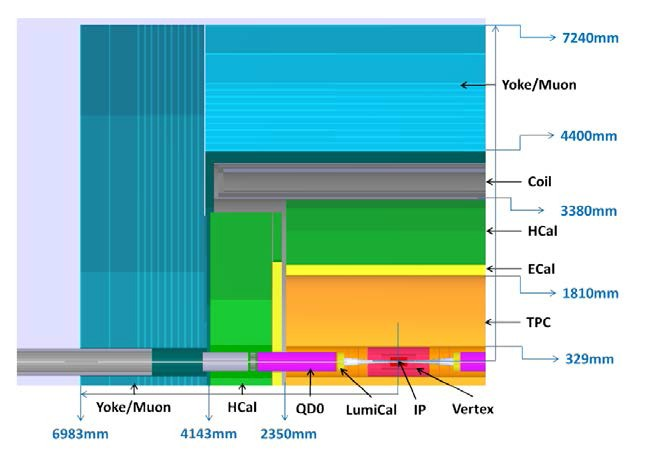}
\centering\includegraphics[width=0.45\textwidth]{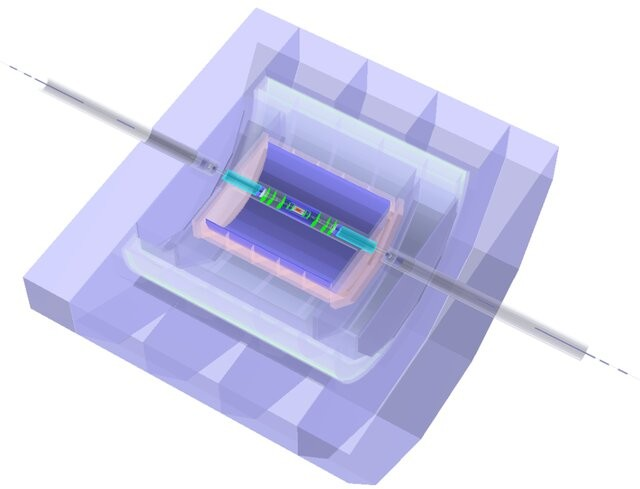}
\caption{The structure (left) and geometry (right) of the baseline CEPC detector design.  From inner to outer, the detector is composed of a
 silicon pixel vertex detector, a silicon inner tracker, a TPC, a silicon external tracker, an ECAL,
 an HCAL, a solenoid of 3 Tesla, and a return yoke incorporating with a muon detector.}
\label{CEPC_structure}
\end{figure}

The baseline detector for CEPC~\cite{CEPCStudyGroup:2018ghi} is designed for a Higgs and high-luminosity $Z$ factory. It features a multi-layer cylindrical architecture as shown in Fig.~\ref{CEPC_structure}~\cite{Sun:2023lut,Liang:2018mst}.
The innermost part is the tracking system, which comprises a silicon pixel vertex detector surrounding the collision region, a silicon inner tracker, and a Time Projection Chamber (TPC) encircled by a silicon external tracker. 
Operating at atmospheric pressure, the TPC~\cite{CEPCStudyGroup:2018ghi} features 220 radial layers and is filled with a gas mixture, offering excellent position information recognition due to its high density of space points.
The cylindrical walls of TPC form the field cage, ensuring a highly homogeneous electric field within the gas mixture for precise $dE/dx$ measurements. The excellent spatial resolution of the silicon trackers, combined with the TPC's $dE/dx$ capabilities, contributes significantly to particle identification (PID).

Surrounding the tracking system is the calorimetry system. A silicon-tungsten sampling electromagnetic calorimeter (Si-W ECAL) precisely measures electromagnetic showers, while a steel-glass resistive plate chambers sampling hadronic calorimeter (SDHCAL) measures hadronic energy deposits. The Si-W ECAL also provides time-of-flight (TOF) information with a time resolution of $50{\mathrm{\,ps}}$~\cite{Zhu:2022hyy}, further enhancing PID capabilities. Furthermore, a 3 Tesla superconducting solenoid is integrated into the design to ensure precise momentum measurement of charged particles, and a flux return yoke incorporating a muon detector in the outermost layer. 
This design follows the particle-flow principle~\cite{Ruan:2013rkk}, optimizing the measurement of final state particles in their most appropriate subdetectors.  

The decay process $B_s^0 \rightarrow D_s^- K^+$ involves the production of a $D_s^-$ meson, which subsequently decays into three charged particles, and an associated kaon. Among the three main decay channels of the $D_s^-$ meson, $D_s^- \rightarrow K^-K^+\pi^-$, $D_s^- \rightarrow K^-\pi^+\pi^-$, and $D_s^- \rightarrow \pi^-\pi^+\pi^-$, only the dominant decay mode, $D_s^- \rightarrow K^-K^+\pi^-$, is simulated. The branching ratios for these decay channels are summarized in Tab.~\ref{tab:branch_ratios}.

\begin{table}[htb]
\begin{center}
\caption{Branching ratios of various decay channels relevant to the \(B_s^0 \to D_s^\mp K^\pm\) analysis, including the dominant decay \(D_s^- \rightarrow K^-K^+\pi^-\) and other significant subchannels.}
\vspace{0.5em}
 \begin{tabular}{c c  }
 \hline\hline
   Branching ratios & Values\\ [0.5ex]
 \hline
${ {\cal B}}(B_s^0 \to D_s^\mp K^\pm )$ & $(2.25 \pm 0.12)\times 10^{-4}$ \cite{PDG2024} \\
\hline
${ {\cal B}} (D_s^- \rightarrow K^-K^+\pi^-)$  & $(5.37 \pm 0.10)\times 10^{-2}$  \cite{PDG2024}  \\
${ {\cal B}}(D_s^- \rightarrow K^-\pi^+\pi^-)$ & $(6.20\pm 0.19)\times 10^{-3}$   \cite{PDG2024} \\
${ {\cal B}}(D_s^- \rightarrow \pi^-\pi^+\pi^-)$ & $(1.08\pm 0.04)\times 10^{-2}$ \cite{PDG2024} \\
\hline
\end{tabular}
\label{tab:branch_ratios}
\end{center}
\end{table}

A Monte Carlo signal sample is generated using the WHIZARD event generator~\cite{Kilian:2007gr} to investigate the geometric acceptance and reconstruction efficiency. The simulation produces ${\totBsEvents}$ $Z\rightarrow b\overline{b} \rightarrow B_s^0X$ events, where $B_s^0$ mesons are exclusively decayed through the channel $B_s^0 \rightarrow D_s^-(\rightarrow K^-K^+\pi^-)K^+$ using PYTHIA8~\cite{Bierlich:2022pfr} with parameter values specified in Tab.~\ref{tab:parameters}. The number of simulated events corresponds to 6.9\% of the expected yield for this specific subchannel ($D_s^- \rightarrow K^-K^+\pi^-$), and 5.3\% of the combined expected yield for all the three $D_s^-$ decay subchannels, under 4.1 Tera-$Z$ boson equivalent data. The simulation methodology employs an existing $B_s^0$ event sample containing multiple decay final states. For each event, a decay channel is randomly selected from the $B_s^0 \rightarrow D_s^\mp K^\pm$ modes, with the initial flavor state and proper lifetime assigned according to the decay rate equations (Eq.~\ref{eq:ideal_decay}). Subsequent adjustments are made to the decay products to ensure physics consistency. Particle transportation through detector materials is simulated using the MokkaC toolkit~\cite{GEANT4:2002zbu}, which implements the GEANT4 framework for tracking particle interactions. 

With the same generator and simulation tool, inclusive samples of $Z\rightarrow q\overline{q}$ events are also generated to estimate the mass and time distributions of the background. Due to the limitation of computing resources, it is not possible to produce as many signal and background events as expected at a Tera-$Z$ factory. The methodology of using a small background sample to generate the background shape is described in Sec.~\ref{sec:inclusive_bkg}, and the full-statistics projection of the $\gamma$ precision is described in Sec.~\ref{sec:fullstats}.

\begin{table}[htb]
\begin{center}
\caption{The values of parameters used in the Monte Carlo signal sample generation for the \(B_s^0 \to D_s^\mp K^\pm\) decays.}
 \vspace{0.5em}
 \begin{tabular}{c c  }
 \hline\hline
   Parameters & Values\\ [0.5ex]
 \hline
$\tau(B_s^0)=1/\Gamma_s$ & $1.520\pm 0.005 \;[ \mathrm{ps}] $~\cite{HFLAV:2022esi} \\
$\Delta \Gamma_s$ & $ +0.084\pm 0.005\; [\mathrm{ps}^{-1}]$ ~\cite{HFLAV:2022esi}\\
\vspace{0.1cm}
$\Delta m_s$& $17.765\pm 0.006\; [\mathrm{ps}^{-1}]$ ~\cite{HFLAV:2022esi}\\
\vspace{0.1cm}
$-2\beta_s$  & $-0.039 \pm 0.016 \;[\mathrm{rad}]$ ~\cite{silva20242023updateextractionckm}    \\
\vspace{0.1cm}
$r_{D_sK}$ & $0.318^{+0.035}_{-0.033}$ ~\cite{LHCb:2024xyw} \\
\vspace{0.1cm}
$\delta$ & $(347.6^{+6.2}_{-6.1})^\circ$ ~\cite{LHCb:2024xyw}\\
\vspace{0.1cm}
$\gamma$ & $(66.2^{+3.4}_{-3.6})^\circ$  ~\cite{HFLAV:2022esi} \\
\hline
\end{tabular}
\label{tab:parameters}
\end{center}
\end{table}

\subsection{Reconstruction and selection}
\label{sec:selection}
The analysis begins with the reconstruction of the decay candidates. For tracking, we simply employ a matching algorithm to correlate reconstructed tracks with Monte Carlo particles, storing suitable matches in linked maps. The LCFIPlus vertex fitter~\cite{Suehara:2015ura} performs sequential vertexing: first reconstructing the \( D_s^- \to K^-K^+\pi^- \) vertex (Tertiary Vertex, TV) with a mass-constrained kinematic fit, then creating a \( D_s^- \) pseudo-track for \( B_s^0 \to D_s^-K^+ \) vertex (Secondary Vertex, SV) fitting with the additional kaon.

\( D_s^- \) candidates are formed from all \(\{\pi, K, K\}\) combinations with correct charge, selecting the candidate whose mass lies within \([ \DsMassToleLow, \DsMassToleHigh ]\)~GeV/\(c^2\) and is closest to the PDG mass of \( D_s^-\), \( \DsPDGMass \)~GeV/\(c^2\)~\cite{PDG2024}. This \( D_s^- \) is paired with a charge-opposite kaon (not from the \( D_s^- \) decay) to form \( B_s^0 \) candidates, choosing the combination with mass nearest the PDG mass of \( B_s^0\), \( \BsPDGMass \)~GeV/\(c^2\)~\cite{PDG2024}, within \([ \BsMassToleLow, \BsMassToleHigh ]\)~GeV/\(c^2\).

To suppress background contributions, we implement a series of cuts based on critical observables associated with the reconstructed vertices. Specifically, we require the chi-squared value of the TV fit to satisfy \(\chi^2_{\text{TV}} < 9\), ensuring a good quality fit. Additionally, we set a minimum distance of $0.5{\mathrm{\,mm}}$ from the TV to the collision point to indicate significant displacement, while the distance from the SV to the collision point must exceed $0.1{\mathrm{\,mm}}$. By combining these invariant mass and vertex cuts, we achieve an acceptance $\times$ efficiency of \(\epsilon = 80\%\) for the signal.

\subsection{Particle identification (PID)}
Sufficient particle identification (PID) efficiency is necessary for flavor physics studies. An effective PID not only minimizes the combinatorial background arising from misidentified particles, but also enables the distinction between decays that exhibit similar topologies in their final states~\cite{Ai:2024nmn}. 
A significant challenge lies in reducing the substantial number of charged pions that are misidentified as charged kaons in our case.
The CEPC Conceptual Design Report (CDR) ~\cite{CEPCStudyGroup:2018ghi} presents a $K/\pi$ separation better than $2\sigma$ over the momentum range up to 20~GeV by effectively combining time-of-flight (TOF) and $dE/dx$ information.
Further optimization studies~\cite{Zhu:2022hyy} have enhanced this performance to $3\sigma$, with $dE/dx$ precision reaching $3\%$ alongside a 50 ps TOF resolution.
Recently, a one-to-one correspondence reconstruction between visible final state particles and reconstructed particles was developed, which would upgrade the identification efficiency of $K^{\pm}$ to $98.5\%$, while for $\pi^{\pm}$ to $97.7\%$~\cite{Wang:2024eji}.

\subsection{Flavor tagging}
\label{sec:tagging}
It is crucial for the $B_s^0 \to D_s^\mp K^\pm$ analysis to identify the initial flavor of the $B_s^0$ meson, which can be performed using different flavor tagging algorithms. The output of these algorithms provides a tag decision \( d = + \), \( - \), or \( 0 \) for a \( B_s^0 \), \( \overline{B}_s^0 \), or an untagged candidate, respectively. All reconstructed \( B_s^0 \) meson decays can be categorized into correctly tagged (R), incorrectly tagged (W), and untagged (U) events. The fraction of events with a non-zero tagging decision is called the tagging efficiency, calculated as \( \epsilon_{\mathrm{tag}} = \frac{R + W}{R + W + U} \). The fraction of events that resulted in a wrong decision is referred to as the wrong tag fraction or mis-tag, which is calculated as \( \omega = \frac{W}{R + W} \). The statistical uncertainty on the measured \( C\!P \) asymmetries is directly related to the effective tagging efficiency, or tagging power, defined as \( \epsilon_{\mathrm{eff}} =  \epsilon_{\mathrm{tag}} (1 - 2\omega)^2 \).

\begin{figure}[htb]
\centering
\centering\includegraphics[width=0.8\textwidth]{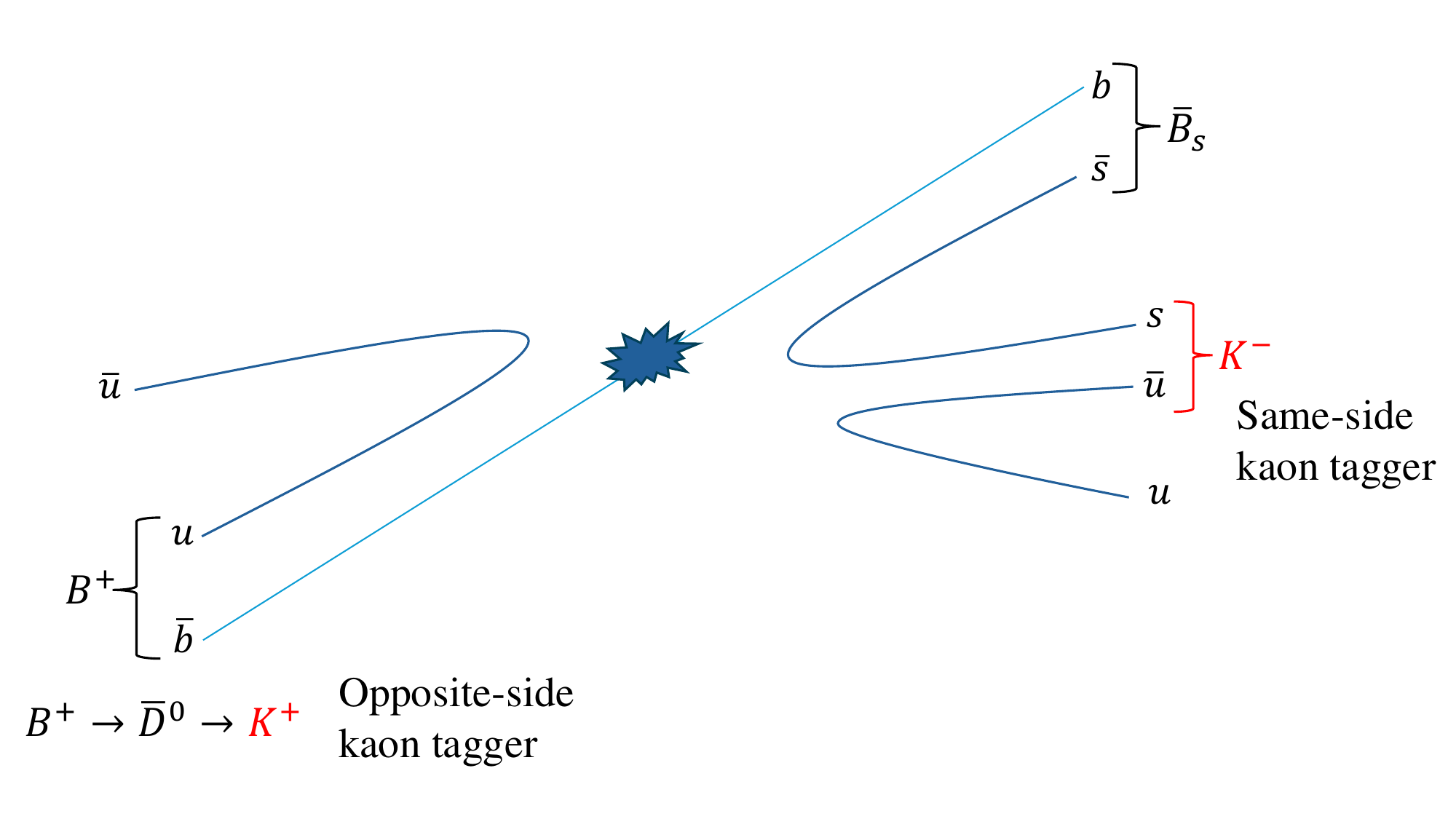}
\caption{Schematic representation of flavor tagging in \( B_s^0 \) decays. The same-side kaon (SSK) arises from fragmentation near the \( B_s^0 \) production vertex, while the opposite-side kaon (OSK) stems from the hadronization and decay chain of the companion \( \overline{b} \) quark. Both kaon types exhibit high energy.}
\label{Bs_decay}
\end{figure}

 Fig.~\ref{Bs_decay} illustrates a flavor tagging mechanism in \( B_s^0 \) decays. In this scenario, a same-side kaon (SSK), such as an accompanying \( K^- \), originates from \( s \)-quark fragmentation and shares the production vertex with the \( \overline{B}_s^0 \). The complementary \( \overline{b} \) quark from the initial quark pair hadronizes into a \( B^+ \), which subsequently decays into a high-energy opposite-side kaon (OSK, e.g., \( K^+ \)) through \( b \to c \to s \) transitions. Both SSKs and OSKs are characterized by significant energy deposition in the detector.

This tagging algorithm prioritizes the leading kaon, defined as the highest-energy kaon that is either in the same direction or in the opposite direction to the \(B_s^0\) flight direction. For SSK identification, the kaon must originate from the \( B_s^0 \) production vertex and lie within a cone of \( 90^\circ \) around the \( B_s^0 \) direction. The kaon charge directly correlates with the initial \( B_s^0 \) flavor: a positively charged kaon indicates a \( B_s^0 \), while a negative charge corresponds to a \( \overline{B}_s^0 \). In contrast, OSKs are identified by their angular separation (\( >90^\circ \)) from the \( B_s^0 \) direction and do not require vertex association with the \( B_s^0 \). 

From truth level events, we got the tagging efficiency results listed in Tab.~\ref{tab:leadingTagging}, which are consistent with the results from~\cite{Li:2022tlo}. The combined result retains the same tag if both algorithms agree, resolves to 0 for conflicting non-zero tags, and adopts the non-zero tag when one algorithm is untagged(d = 0).

\begin{table}[htbp] \begin{center} 
\caption{ Tagging efficiency results from truth-level events for the leading kaon tagging algorithm. The combined tagging outcomes are evaluated, with agreements retained, conflicts resolved to 0, and non-zero tags adopted when one algorithm is untagged.} 
 \vspace{0.5em}
\begin{tabular}{c c c c} \hline\hline Methods & $\epsilon_\mathrm{tag}$ & $\omega$ & $\epsilon_\mathrm{eff}$ \\ [0.5ex] \hline Leading SSK (\%) & 57.45 & 19.96 & 20.73 \\ Leading OSK (\%) & 72.41 & 31.36 & 10.06 \\  Combined (\%) & 72.12 & 21.39 & 23.60 \\ \hline \end{tabular}
\label{tab:leadingTagging}
\end{center}
\end{table}

\subsection{Decay time resolution and acceptance}
\label{sec:decay_resolution}
\begin{figure}[htb]
\centering
\centering\includegraphics[width=0.6\textwidth]{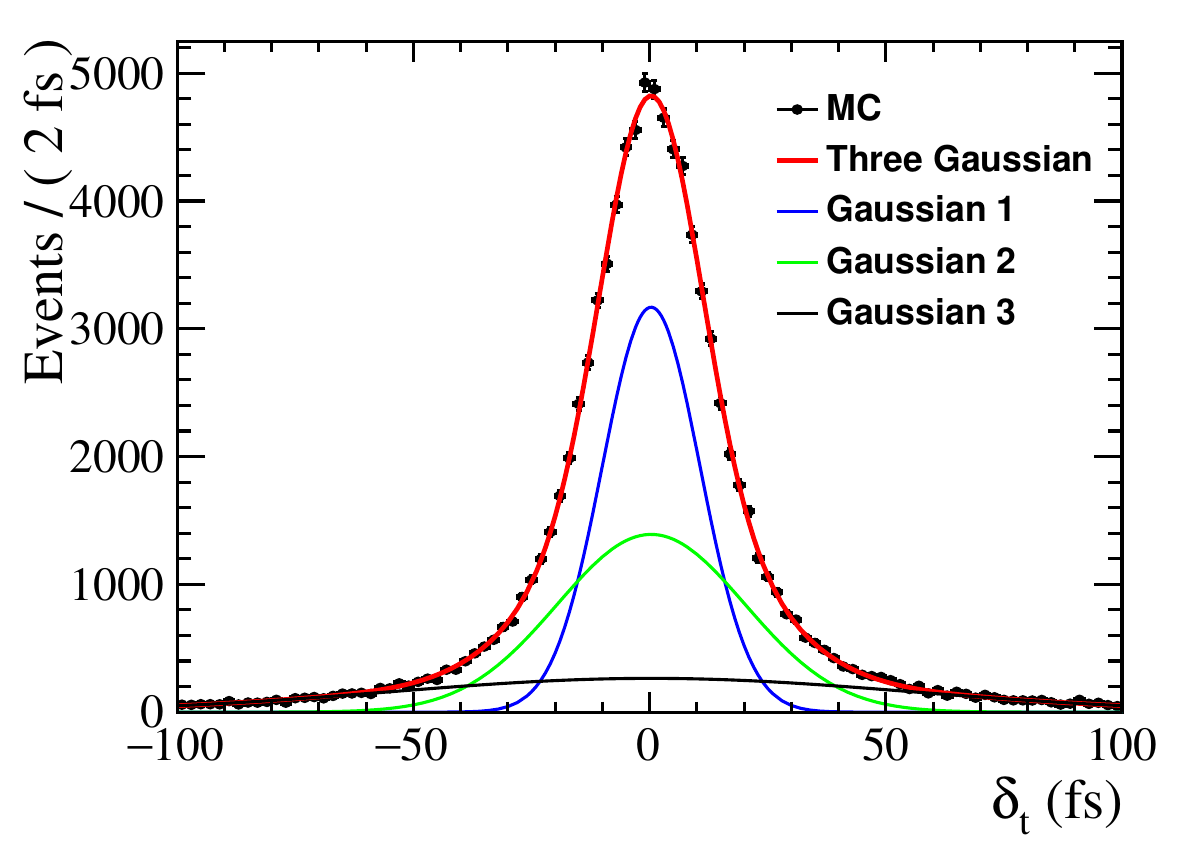}
\caption{Proper time resolution distribution of \( B_s^0 \) candidates, modeled as a mixture of three Gaussian functions achieving an effective resolution of \( \sigma_t = 26\mathrm{\,fs} \). }
\label{time_resolution}
\end{figure}

A dilution of the observable oscillation is caused by the finite decay time resolution of the detector.  Upon reconstructing the \( B_s^0\) decay vertex, the spatial resolution in the \( z \)-direction is found to be less precise than in the \( x \) and \( y \) directions. Therefore, the \( B_s^0\) lifetime can be calculated using the reconstructed vertex position and transverse momentum, \( p_T \):
\begin{eqnarray}
    \tau = \frac{m_s l_{xy}}{p_T},
\end{eqnarray}
where \( \tau \) is the proper time of the \( B_s^0\) decay, \( m_s \) is its mass, \( l_{xy} \) is the decay length in the transverse plane. This leads to a time residual distribution, illustrated in Fig.~\ref{time_resolution}. A mixture of three Gaussian functions, \( Res(t, \sigma_1, \sigma_2, \sigma_3) \),  sharing the same mean value models this distribution as follows:
\begin{eqnarray}
    Res(t, \sigma_1, \sigma_2, \sigma_3) = f_1 G(t, \sigma_1) + f_2 G(t, \sigma_2) + (1-f_1-f_2) G(t, \sigma_3),
\end{eqnarray}
where \( f_{1,2,3} \) are the fractions, and \( \sigma_{1,2,3} \) denote the widths of the three Gaussian functions $G(t, \sigma_{1,2,3})$. This distribution is obtained from simulation after full \( B_s^0 \) reconstruction, and the effective resolution is extracted from it.

In this analysis, the nominal value of the decay time error is derived from a method employed in several LHCb time-dependent studies, such as \cite{LHCb:2024xyw}, which was first described in \cite{Aaij:2015ijn}. According to this approach, the resolutions of the three Gaussian functions are combined into an effective resolution \( \sigma_t = 26\mathrm{\,fs}\):
\begin{eqnarray}
 \sigma_t = \sqrt{-\frac{2}{\Delta m_s^2}\ln\left( f_1 e^{-\frac{1}{2}\sigma_1^2 \Delta m_s^2 } + f_2 e^{-\frac{1}{2}\sigma_2^2 \Delta m_s^2 } + (1-f_1-f_2) e^{-\frac{1}{2}\sigma_3^2 \Delta m_s^2 }\right) }.  
\end{eqnarray}

The measured decay times of $B_s^0$ mesons are influenced not only by the detector resolution but also by selection criteria applied during event reconstruction. These combined effects are formally described through a decay time acceptance function $a(t)$, which modifies the theoretical decay time distribution as follows:
\begin{equation} \frac{d\Gamma^\text{acc}(t)}{dt} = \frac{d\Gamma(t)}{dt} \times a(t). \end{equation}

Following established methodologies in analysis~\cite{LHCb:2011aa}, we employ an empirical acceptance function parameterized by $(\alpha, \beta, \xi)$:
\begin{equation} a(t) = \frac{(\alpha t)^\beta}{1+(\alpha t)^\beta}(1-\xi t). \end{equation}

\section{Multivariate Fitting Strategy}
The analysis employs a simultaneous fit to invariant mass and decay time distributions. Following flavor tagging classification, the dataset is partitioned into six subsamples based on $B^0_s$ flavor assignment and decay products. These subsamples can be
categorized as follows:
\begin{eqnarray} 
&B^0_{s,+} \to D_s^+ K^-, \quad B^0_{s,-} \to D_s^+ K^-, \quad B^0_{s,0} \to D_s^+ K^-, \notag\\ &B^0_{s,+} \to D_s^- K^+, \quad B^0_{s,-} \to D_s^- K^+, \quad B^0_{s,0} \to D_s^- K^+. \notag\end{eqnarray}

\subsection{Signal parametrization}
For the signal mass shape parametrization, a double-sided Crystal Ball function~\cite{Skwarnicki:1986xj} is employed with a symmetric Gaussian core and symmetric tails. It is reasonable to assume that all six subsamples share the same mass shape.

The signal decay time distributions of the six subsamples are more complex. Due to flavor tagging, they are mixtures of the four ideal distributions represented as Eq.~\ref{eq:ideal_decay}. Furthermore, these distributions are distorted by decay time resolution and acceptance. Thus, the final decay time distributions are:
\begin{multline}
\begin{pmatrix} P_{B^0_{s,+} \to D_s^+ K^-}(t) \\ P_{B^0_{s,-} \to D_s^+ K^-}(t)  \\P_{B^0_{s,0} \to D_s^+ K^-}(t) \\ P_{B^0_{s,+} \to D_s^- K^+}(t) \\ P_{B^0_{s,-}\to D_s^- K^+}(t)\\  P_{B^0_{s,0} \to D_s^- K^+}(t) \end{pmatrix} \propto 
 \underbrace{\begin{pmatrix} 1 - \omega & \omega & 0 & 0 \\ \omega & 1 - \omega & 0 & 0 \\ \epsilon_{\mathrm{tag}}^{-1}-1 & \epsilon_{\mathrm{tag}}^{-1}-1 & 0 & 0 \\ 0 & 0 & 1 - \omega & \omega \\ 0 & 0 & \omega & 1 - \omega \\ 0 & 0 & \epsilon_{\mathrm{tag}}^{-1}-1 & \epsilon_{\mathrm{tag}}^{-1}-1 \end{pmatrix}}_{\text{Tagging matrix}} \\
 \cdot \underbrace{\left[ \begin{pmatrix} P_{B_s^0 \rightarrow D_s^+ K^-}(t)\\ P_{\overline{B}_s^0 \rightarrow D_s^+ K^-}(t)  \\P_{B_s^0 \rightarrow D_s^- K^+}(t) \\  P_{\overline{B}_s^0 \rightarrow D_s^- K^+}(t)\end{pmatrix}
 \theta(t) \right]}_{\text{Ideal PDFs}}
\otimes \underbrace{Res(t, \sigma_1, \sigma_2, \sigma_3)\cdot a(t)}_{\text{Detector effects}}.
\label{eq:reco_decay}
\end{multline}

\subsection{Background treatment}
\label{sec:inclusive_bkg}
 It is challenging to generate a sufficient number of fully simulated inclusive background samples of $Z\rightarrow q\overline{q}$ events (hereafter referred to as the $Z\rightarrow q\overline{q}$ sample) to directly extract the background shape. Therefore, an alternative method is developed to estimate both mass and time distributions of background from the limited available $Z\rightarrow q\overline{q}$ samples. An extended background dataset (hereafter referred to as the generated background samples) is then constructed based on the obtained distributions, and used in the final fit.

\begin{figure}[htb]
\centering
\centering\includegraphics[width=0.45\textwidth]{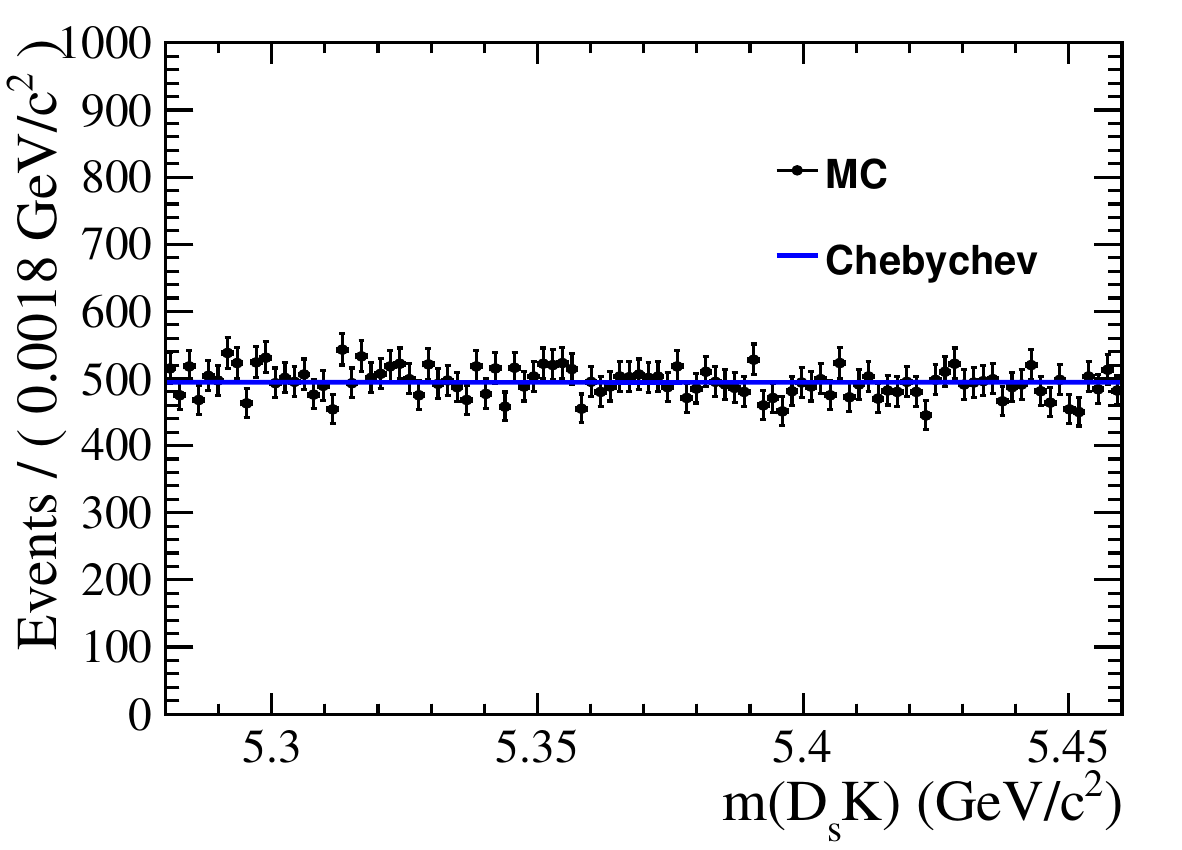}
\centering\includegraphics[width=0.45\textwidth]{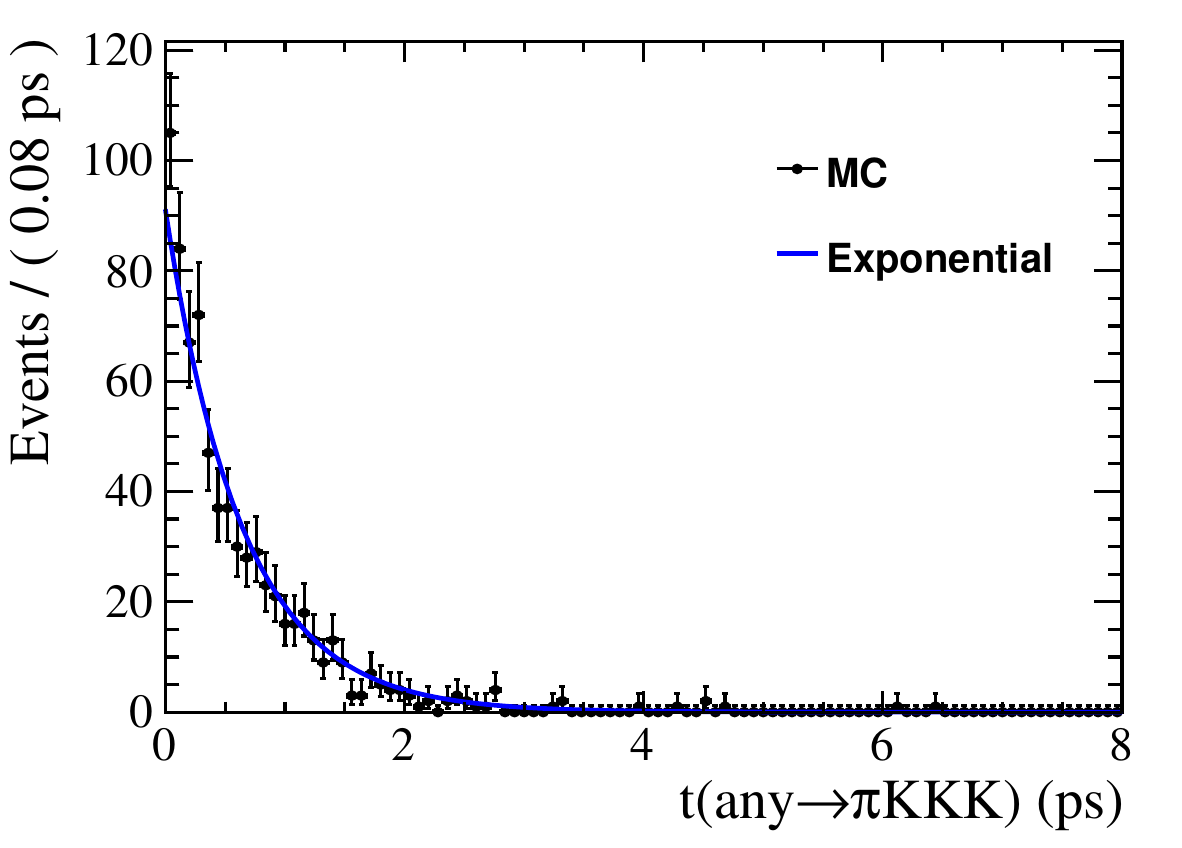}
\caption{Background estimation for fake \( B_s^0 \) candidates. The left plot shows the mass distribution of generated fake candidates, constrained by the vertex distance of kaon and \( D_s^-\) to be less than $0.05{\mathrm{\,mm}}$ and within the \( B_s^0 \) mass window, fitted with a first-order Chebyshev polynomial. The right plot depicts the decay time distribution of the inclusive background, modeled using an exponential function, reflecting the same decay time shape as the signal decay for final states \( \{ \pi, K, K, K \} \).}
\label{inclusive_bkg_shape}
\end{figure}

For the mass side, the four-momentum and vertex distributions of the kaons and the \( D_s^-\) candidates are obtained from the $Z\rightarrow q\overline{q}$ samples.  We assume that the distributions of the kaon and \( D_s^-\) that form a fake \( B_s^0\) candidate reflect similar characteristics. To estimate the background level, we generate a kaon and a \( D_s^-\) to create a fake \( B_s^0\) candidate, ensuring that the distance between their true vertices is less than $0.05{\mathrm{\,mm}}$. Additionally, this fake candidate must fall within the \( B_s^0\) mass window to be relevant for the inclusive mass background shape, as illustrated on the left in Fig.~\ref{inclusive_bkg_shape}. We employ a first-order Chebyshev polynomial to fit this background shape. The probability of reconstructing a fake \( B_s^0\) candidate is estimated to be \( 5 \times 10^{-7} \), and the resulting background statistics are comparable to those of the signal.

For the time side, it is assumed that the inclusive background exhibits the same decay time shape as the true decay time distribution extracted from the $Z\rightarrow q\overline{q}$ samples of particles decaying into the final states \( \{ \pi, K, K, K \} \). These final states correspond to the products of the signal decay. An exponential function is used to model this distribution, as shown on the right in Fig.~\ref{inclusive_bkg_shape}. Additionally, all six subsamples are assumed to share identical background characteristics in terms of both shape and normalization.

\subsection{Simultaneous mass-time fit}
After the analysis of the signal and inclusive background, an unbinned two-dimensional simultaneous maximum likelihood fit is performed. For each subsample, the fit model is given by:
\begin{eqnarray}
 \textit{Model\,}(m,t) = N_s S(m) S(t) + N_b B(m) B(t), 
 \end{eqnarray}
where \( N_s \) represents the signal yields, and \( N_b \) denotes the background yields. As discussed previously, the signal mass model \( S(m) \) follows a Double-Sided Crystal Ball function, and the decay time distribution for the signal, \( S(t) \), refers to the functions outlined in Eq.~\ref{eq:reco_decay}. The background shapes \( B(m) \) and \( B(t) \) are modeled using a Chebyshev polynomial for the mass distribution and a distorted exponential function (an exponential function convolved with the time resolution and multiplied by the time acceptance) for the time distribution, assuming the background shares the same resolution and acceptance as the signal.

\begin{figure}[htb]
\centering
\centering\includegraphics[width=0.45\textwidth]{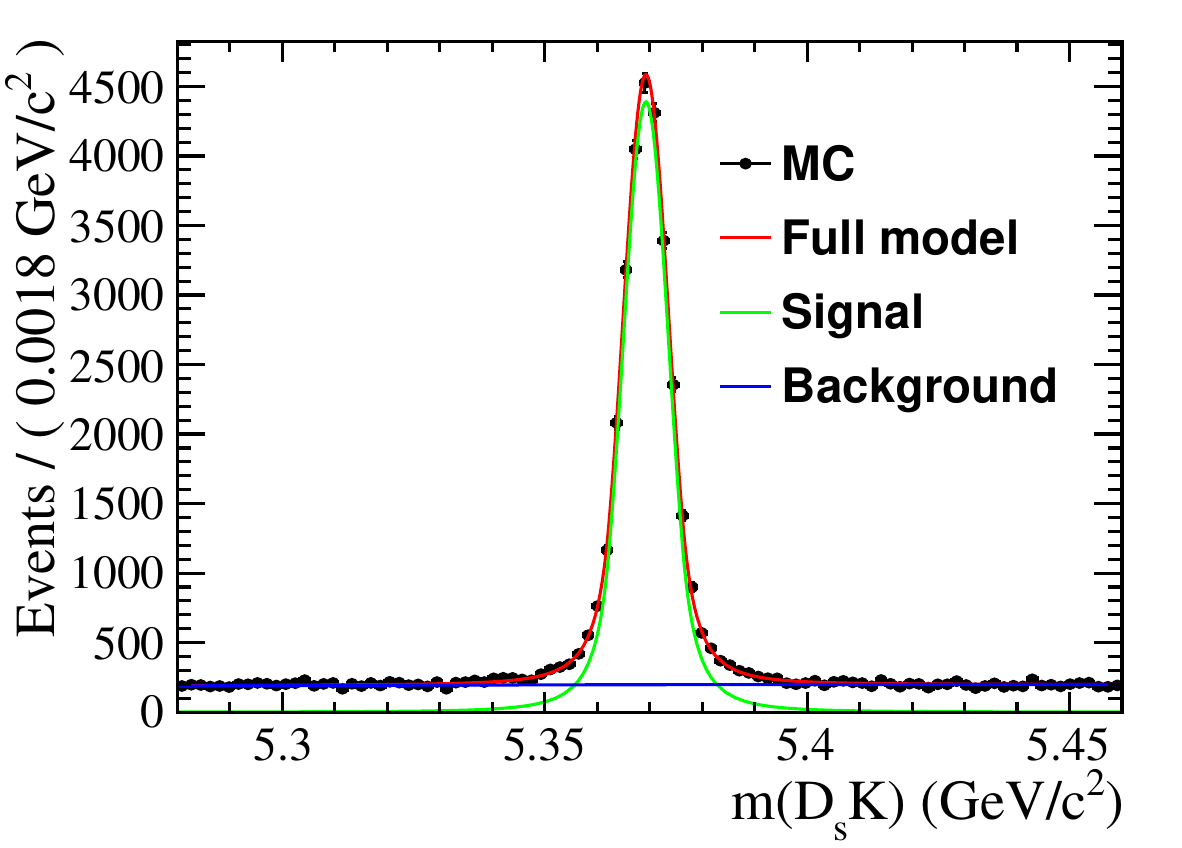}
\put(-155, 100){\fontsize{7}{5}\selectfont $\mathbf{B^0_{s,+} \to D_s^+ K^-}$}
\centering\includegraphics[width=0.45\textwidth]{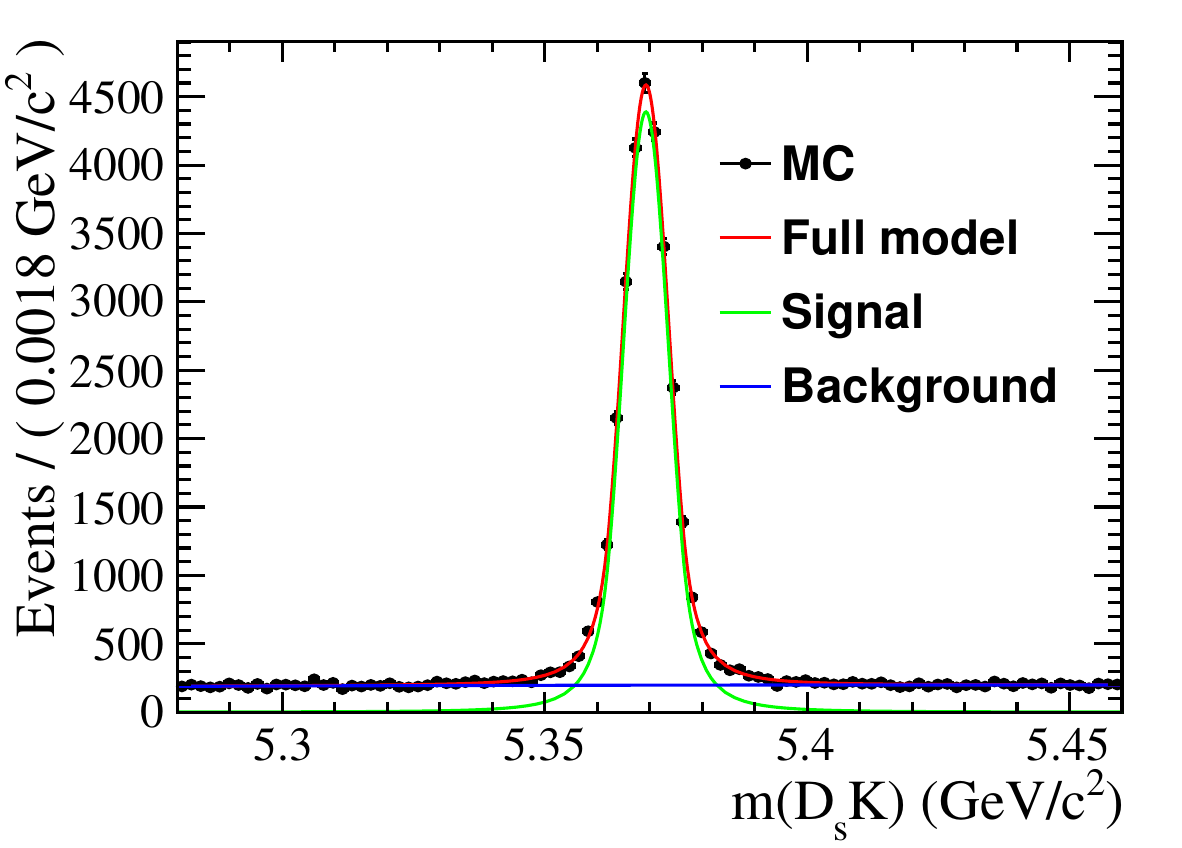}
\put(-155, 100){\fontsize{7}{5}\selectfont $\mathbf{B^0_{s,+} \to D_s^- K^+}$}

\centering\includegraphics[width=0.45\textwidth]{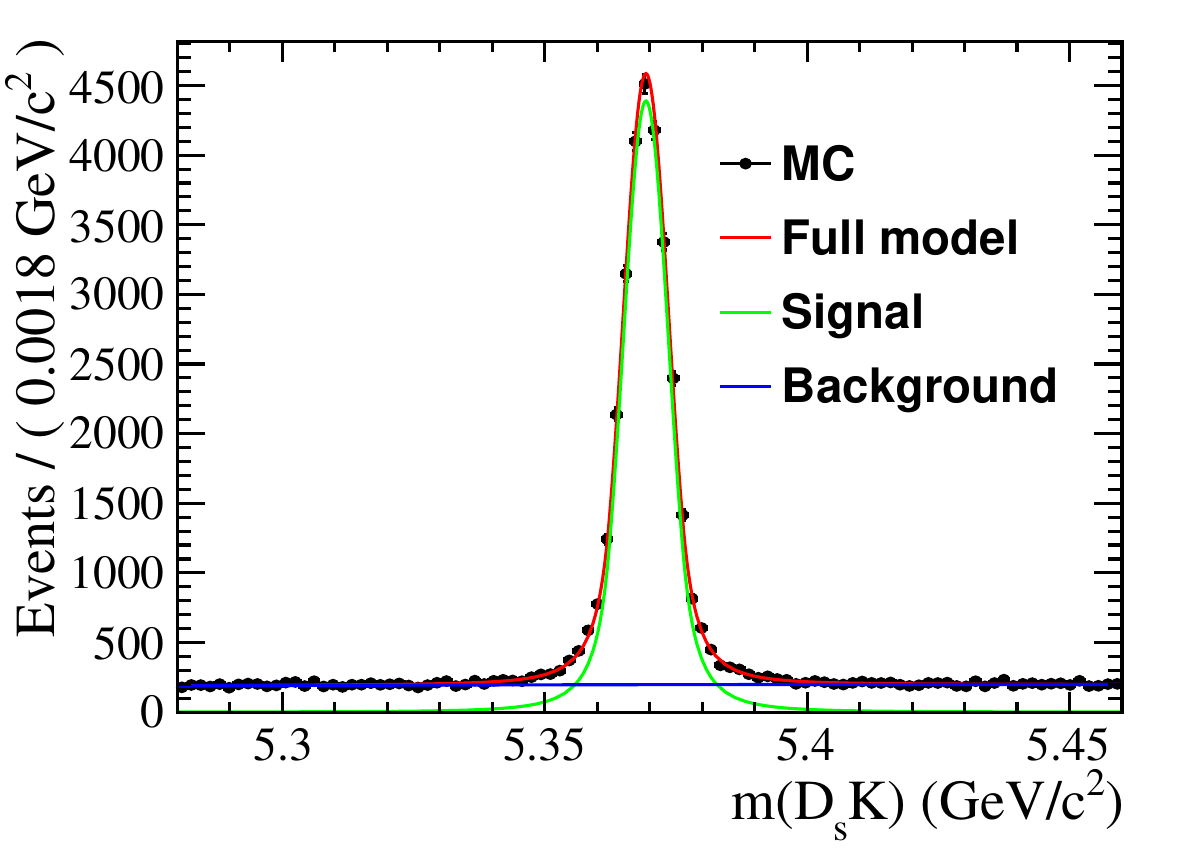}
\put(-155, 100){\fontsize{7}{5}\selectfont $\mathbf{B^0_{s,-} \to D_s^+ K^-}$}
\centering\includegraphics[width=0.45\textwidth]{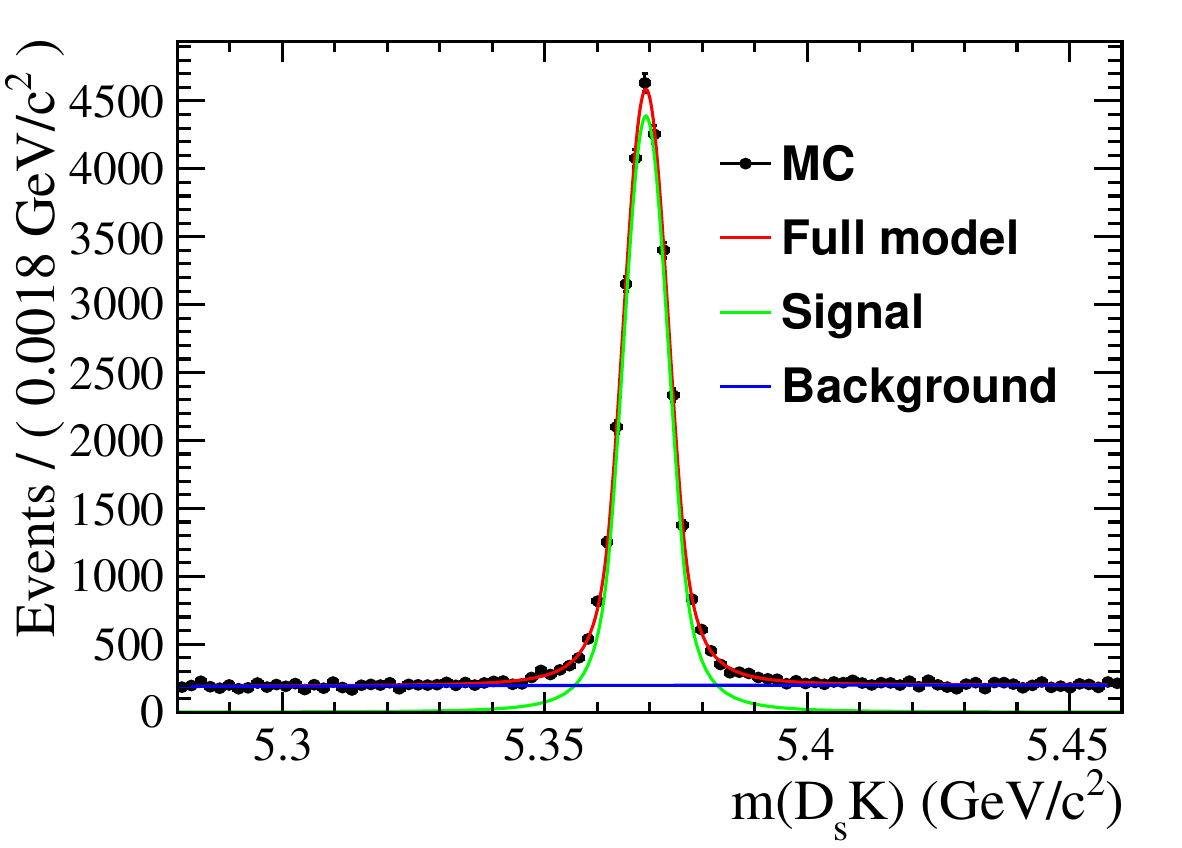}
\put(-155, 100){\fontsize{7}{5}\selectfont $\mathbf{B^0_{s,-} \to D_s^- K^+}$}

\centering\includegraphics[width=0.45\textwidth]{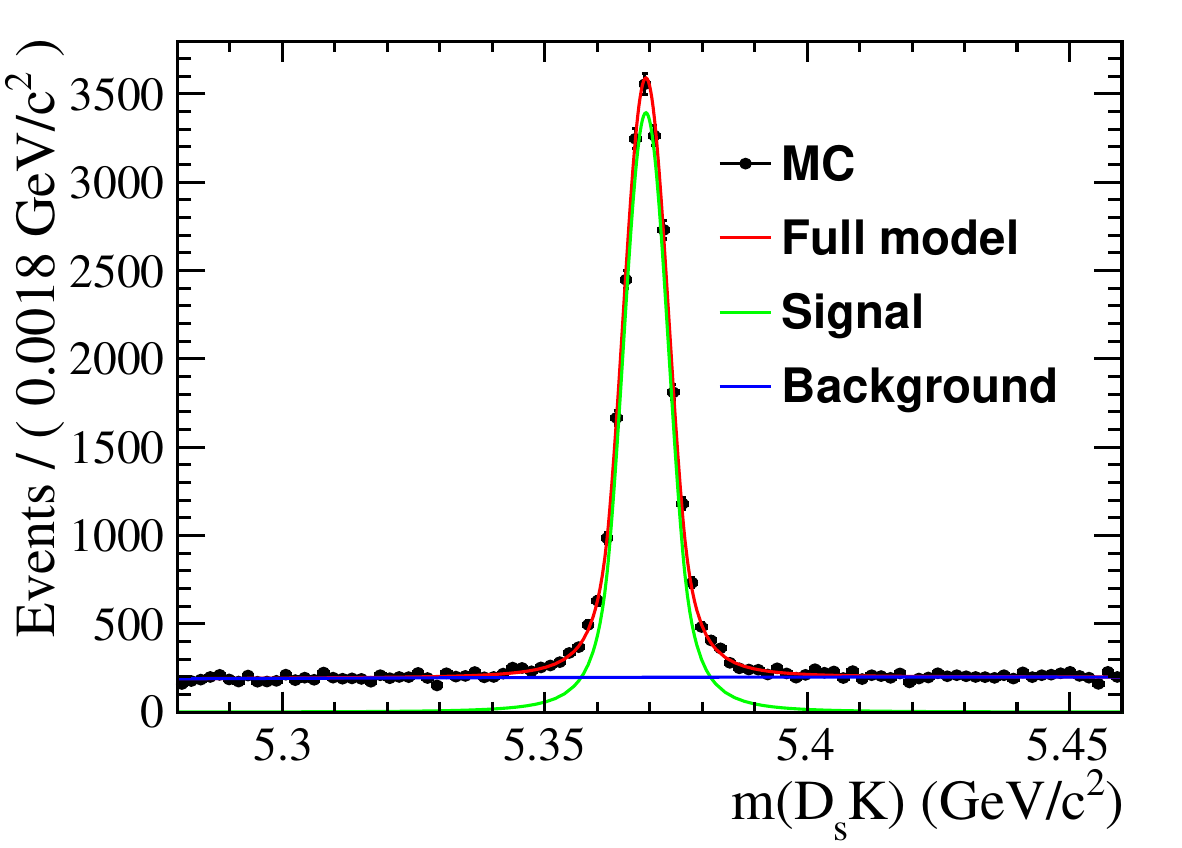}
\put(-155, 100){\fontsize{7}{5}\selectfont $\mathbf{B^0_{s,0} \to D_s^+ K^-}$}
\centering\includegraphics[width=0.45\textwidth]{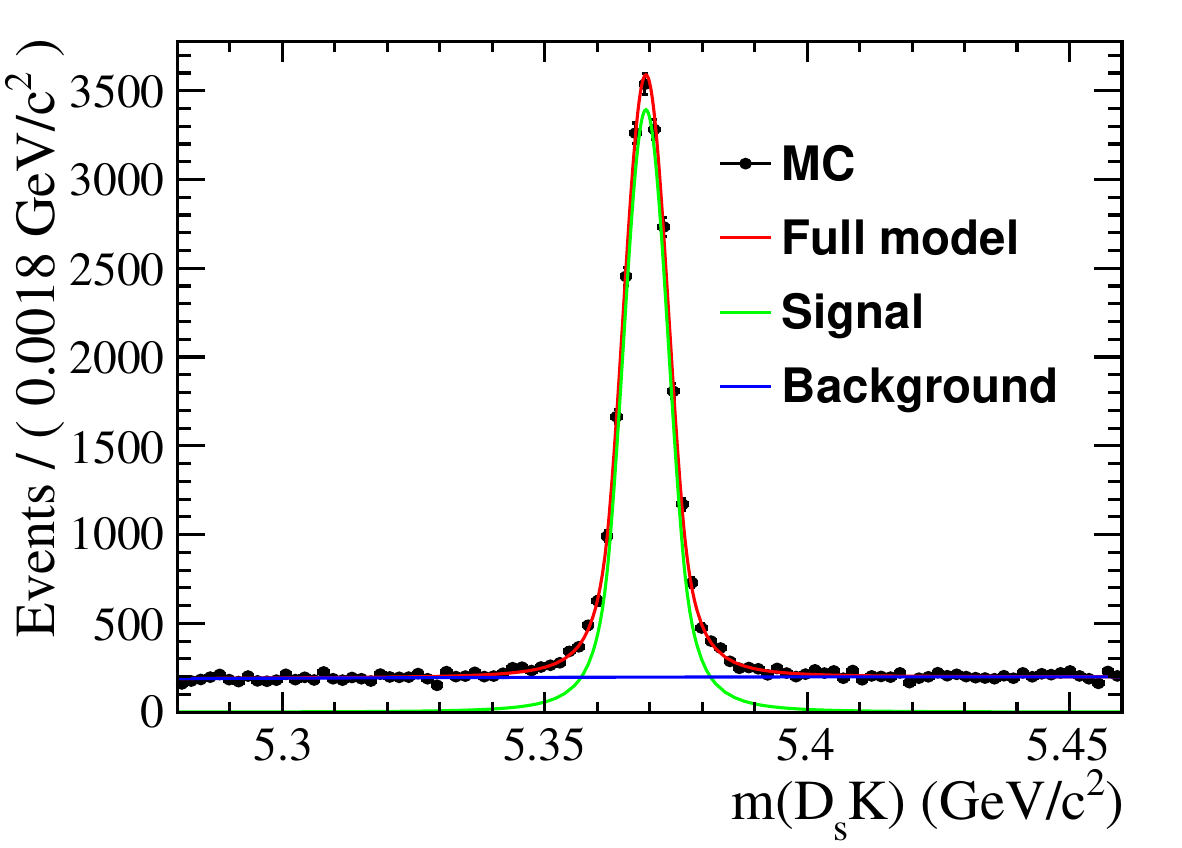}
\put(-155, 100){\fontsize{7}{5}\selectfont $\mathbf{B^0_{s,0} \to D_s^- K^+}$}
\caption{Results from the two-dimensional simultaneous fit showing the signal and background contributions to the mass distribution. The fit model incorporates a Double-Sided Crystal Ball function for the signal mass distribution, while the background is modeled using a Chebyshev polynomial.}
\label{fig:mass_plots}
\end{figure}

\begin{figure}[htb]
\centering
\centering\includegraphics[width=0.45\textwidth]{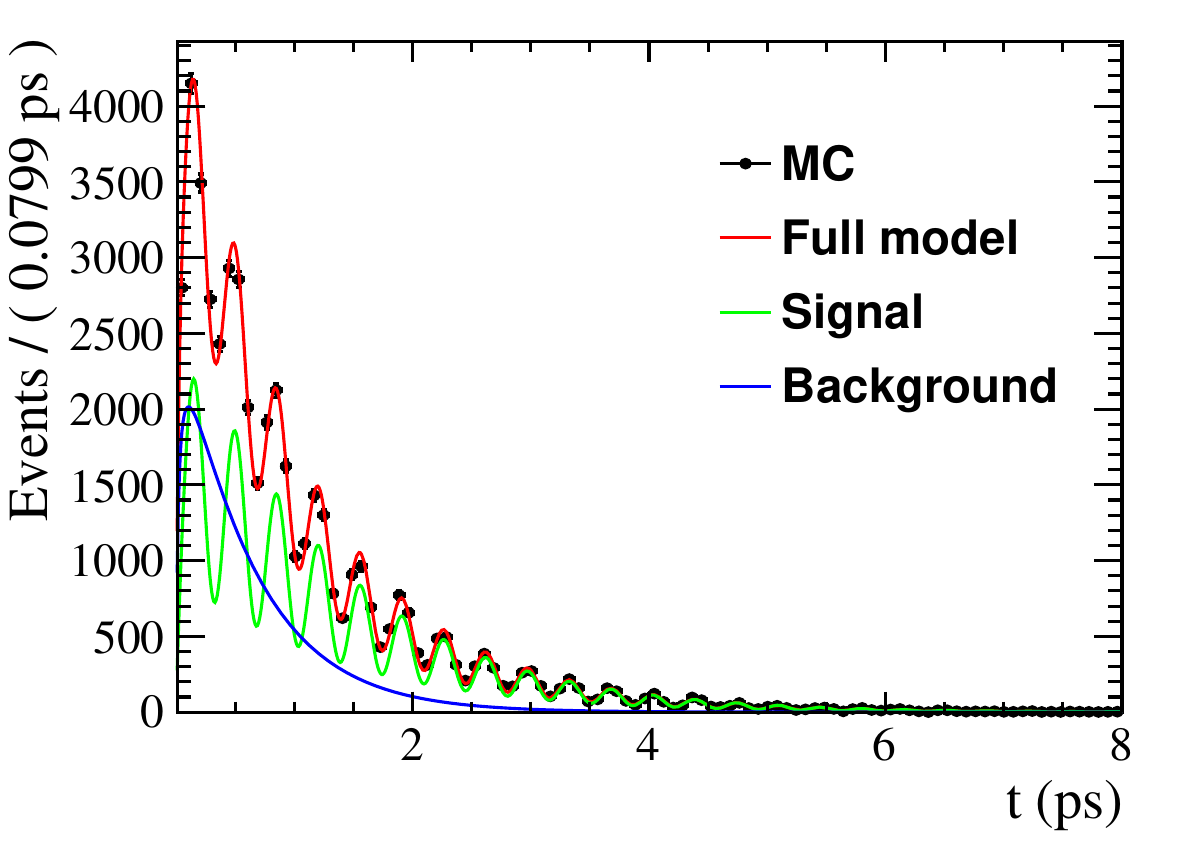}
\put(-140,100){\fontsize{7}{5}\selectfont $\mathbf{B^0_{s,+} \to D_s^+ K^-}$}
\centering\includegraphics[width=0.45\textwidth]{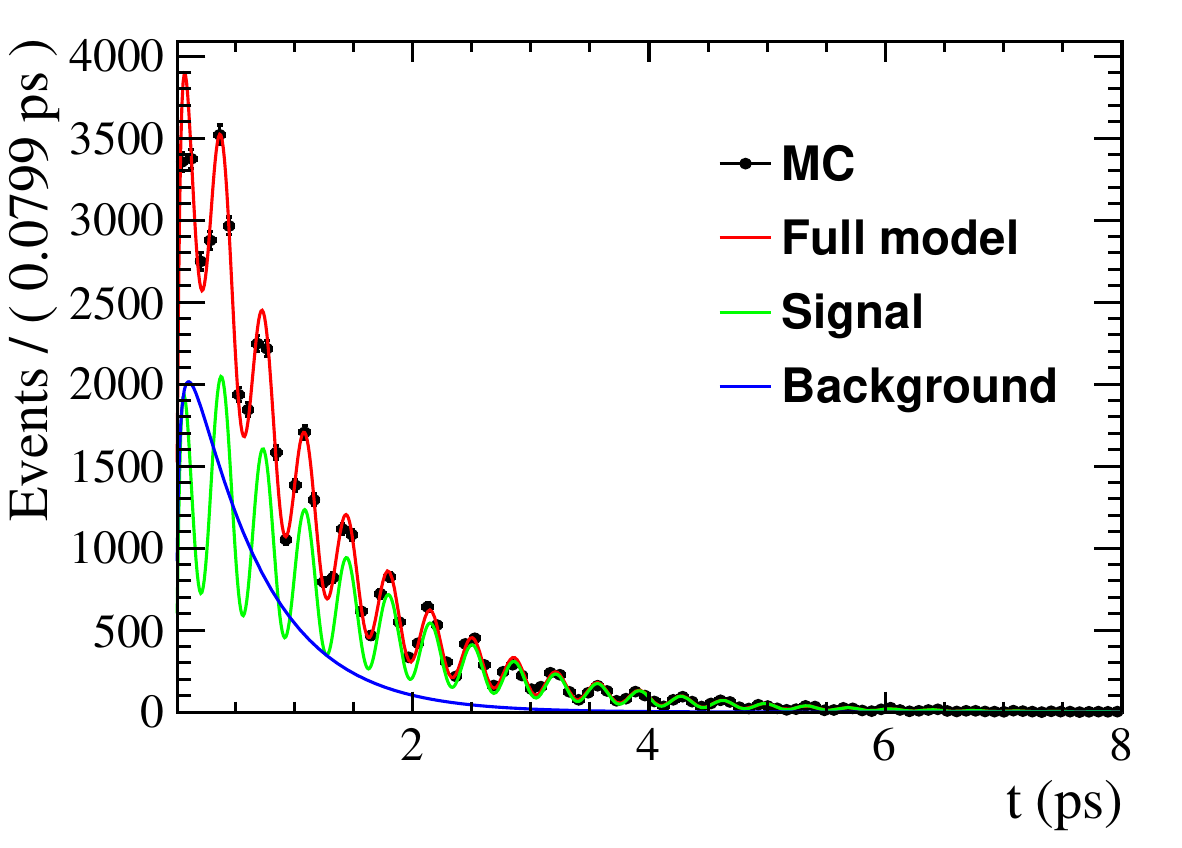}
\put(-140,100){\fontsize{7}{5}\selectfont $\mathbf{B^0_{s,+} \to D_s^- K^+}$}

\centering\includegraphics[width=0.45\textwidth]{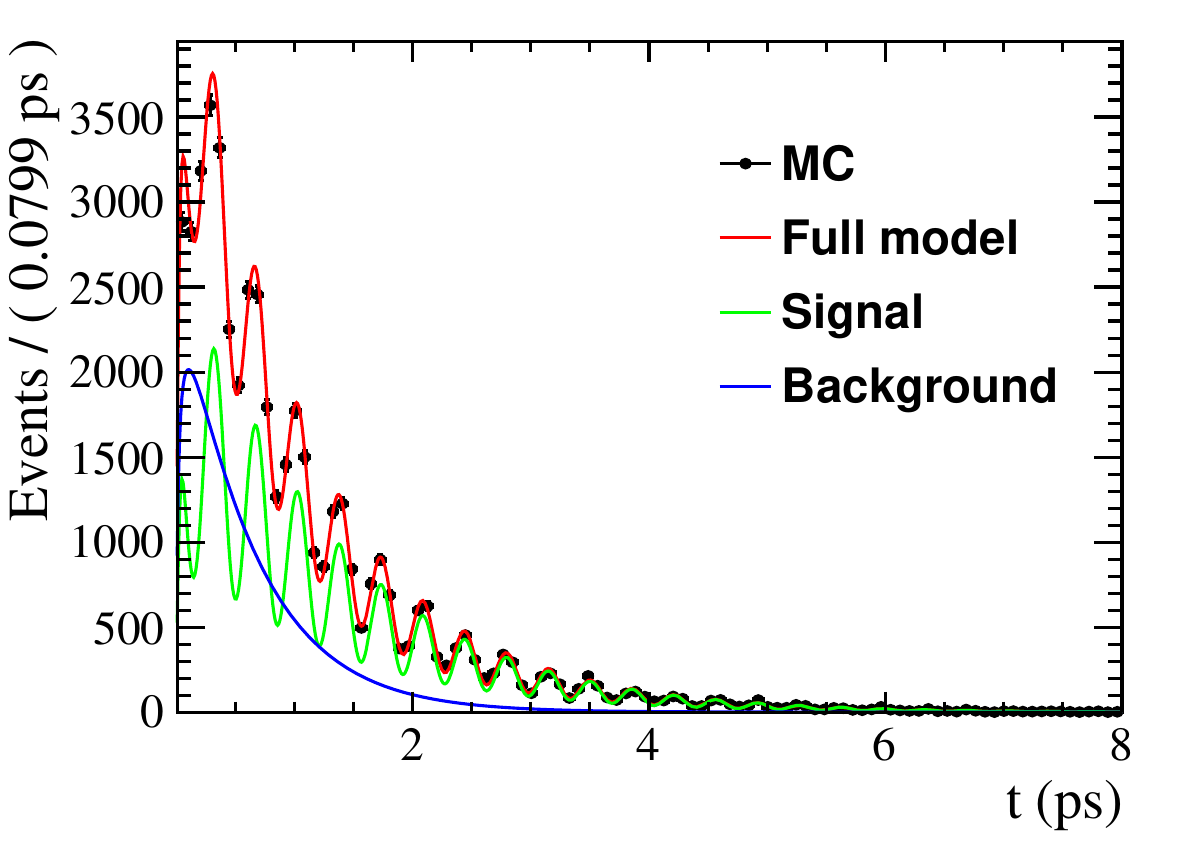}
\put(-140,100){\fontsize{7}{5}\selectfont $\mathbf{B^0_{s,-} \to D_s^+ K^-}$}
\centering\includegraphics[width=0.45\textwidth]{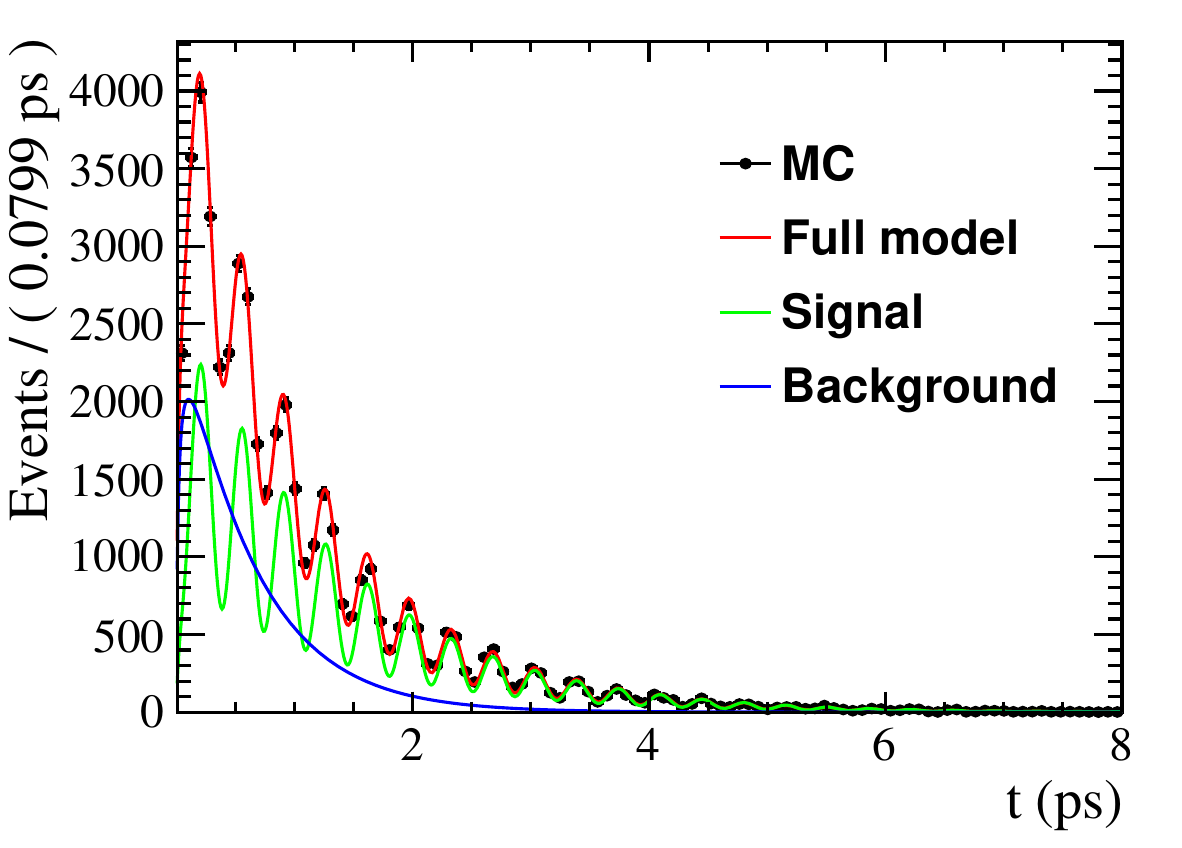}
\put(-140,100){\fontsize{7}{5}\selectfont $\mathbf{B^0_{s,-} \to D_s^- K^+}$}

\centering\includegraphics[width=0.45\textwidth]{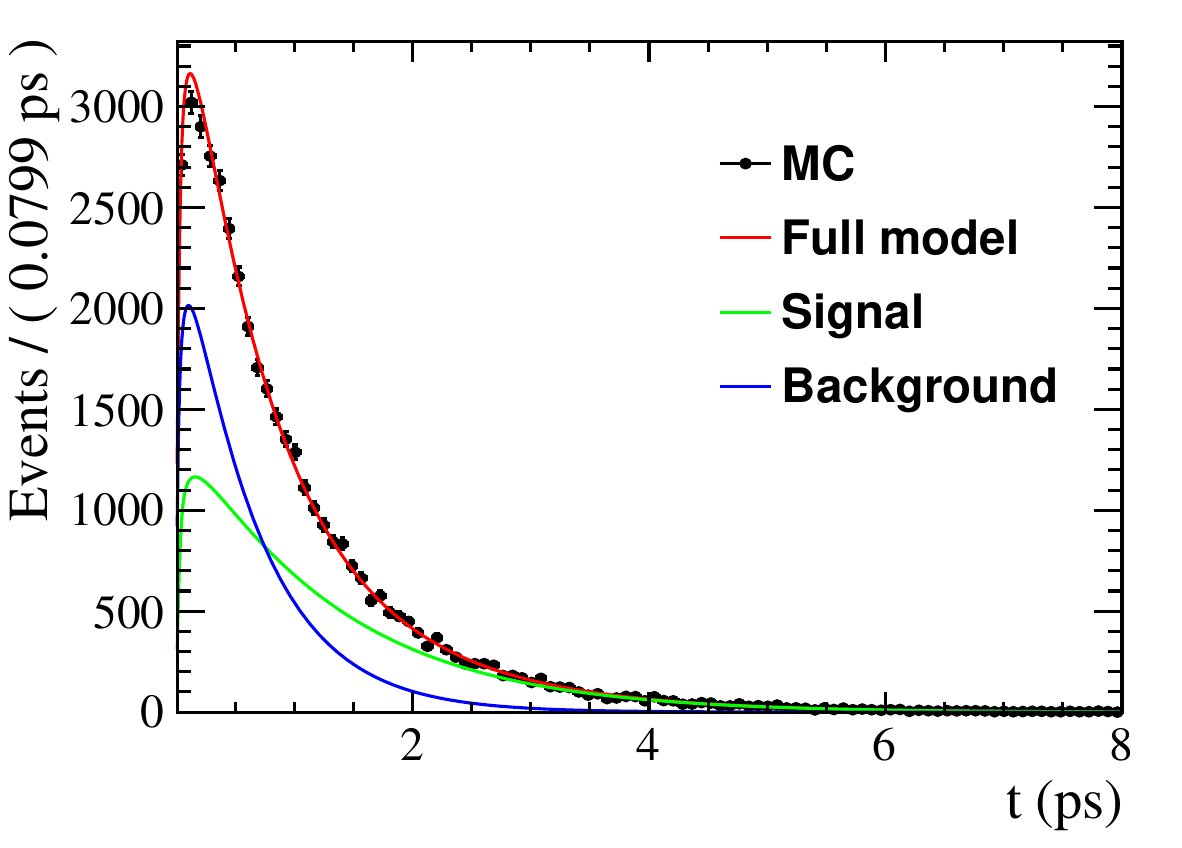}
\put(-140,100){\fontsize{7}{5}\selectfont $\mathbf{B^0_{s,0} \to D_s^+ K^-}$}
\centering\includegraphics[width=0.45\textwidth]{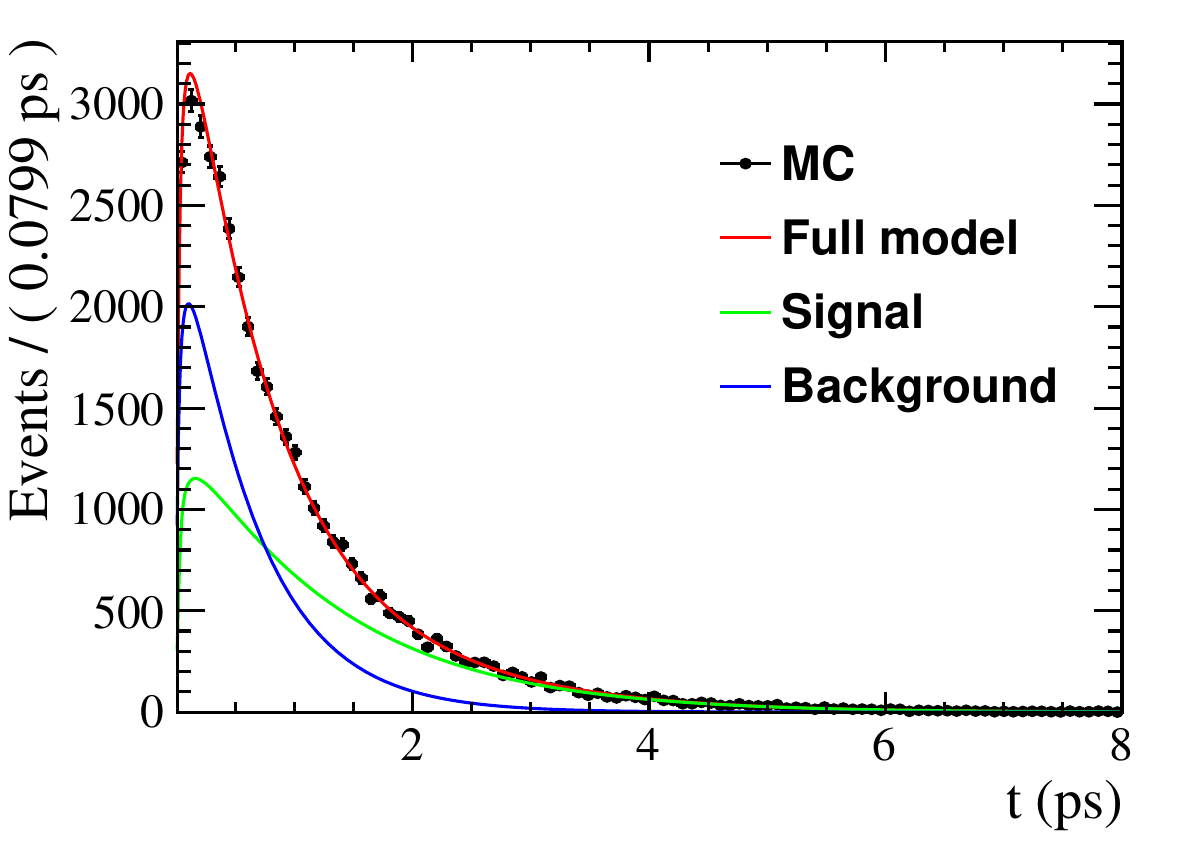}
\put(-140,100){\fontsize{7}{5}\selectfont $\mathbf{B^0_{s,0} \to D_s^- K^+}$}
\caption{Time distribution projections from the two-dimensional simultaneous fit illustrating the signal and background components. The decay time distribution for the signal is derived from the distributions defined in Eq.~\ref{eq:reco_decay}, while the background is modeled with a distorted exponential function.}
\label{fig:time_plots}
\end{figure}

The entire dataset is a mixture of signal and background data. The signal data is reconstructed from the signal samples, while the background data is taken from the generated background samples. These background samples are generated to correspond to the simulated signal statistics. The total background generated is $1.18 \times 10^5$, which is of the same order as the signal. In this fit, the values of \( \Gamma_s \), \( \Delta \Gamma_s \), \( \Delta m_s \), and \( -2\beta_s \) are fixed as shown in Tab.~\ref{tab:parameters}. After conducting the two-dimensional fit, the mass projection and time projection plots are presented in Fig.~\ref{fig:mass_plots} and Fig.~\ref{fig:time_plots}, respectively.

\section{Results and Discussion}
Following the simultaneous fit of invariant mass and decay time of $B_s^0$, we obtain the values and associated uncertainties of parameters of interest summarized in Tab.~\ref{tab:fit_results_real_v1}, where the angles are given modulo $180^\circ$. Notably, the resolution of $\gamma$ is determined to be \(3.01^\circ\) based on ${\totBsEvents}$ simulated $Z \to b\overline{b} \to B_s^0 X$ events, where $B_s^0$ mesons decay through $B_s^0 \to D_s^-(\to K^-K^+\pi^-)K^+$. This corresponds to 5.3\% of the total expected yield for all three $D_s^-$ decay subchannels, with 4.1 Tera-$Z$ boson equivalent data.

\begin{table}[htb]
\renewcommand{\arraystretch}{1.3}
\begin{center}
\caption{Summary of fit results for key parameters obtained from the simultaneous analysis of invariant mass and decay time distributions. The angles are presented modulo \(180^\circ\).}
 \vspace{0.5em}
 \begin{tabular}{c c  }
 \hline\hline
   parameters & values \\ [0.5ex]
 \hline
$r_{D_sK}$ & $0.314  \pm 0.007 $  \\
$\delta$ & $(349.64 \pm 2.39)^{\circ} $ \\
$\gamma$ & $(66.43 \pm \gammaAnaErr)^{\circ}$ \\
[1ex]
\hline
\end{tabular}
\label{tab:fit_results_real_v1}
\end{center}
\end{table}

\subsection{Key factors: tagging power, time resolution}
\begin{figure}[htb]
\centering
\centering\includegraphics[width=0.45\textwidth]{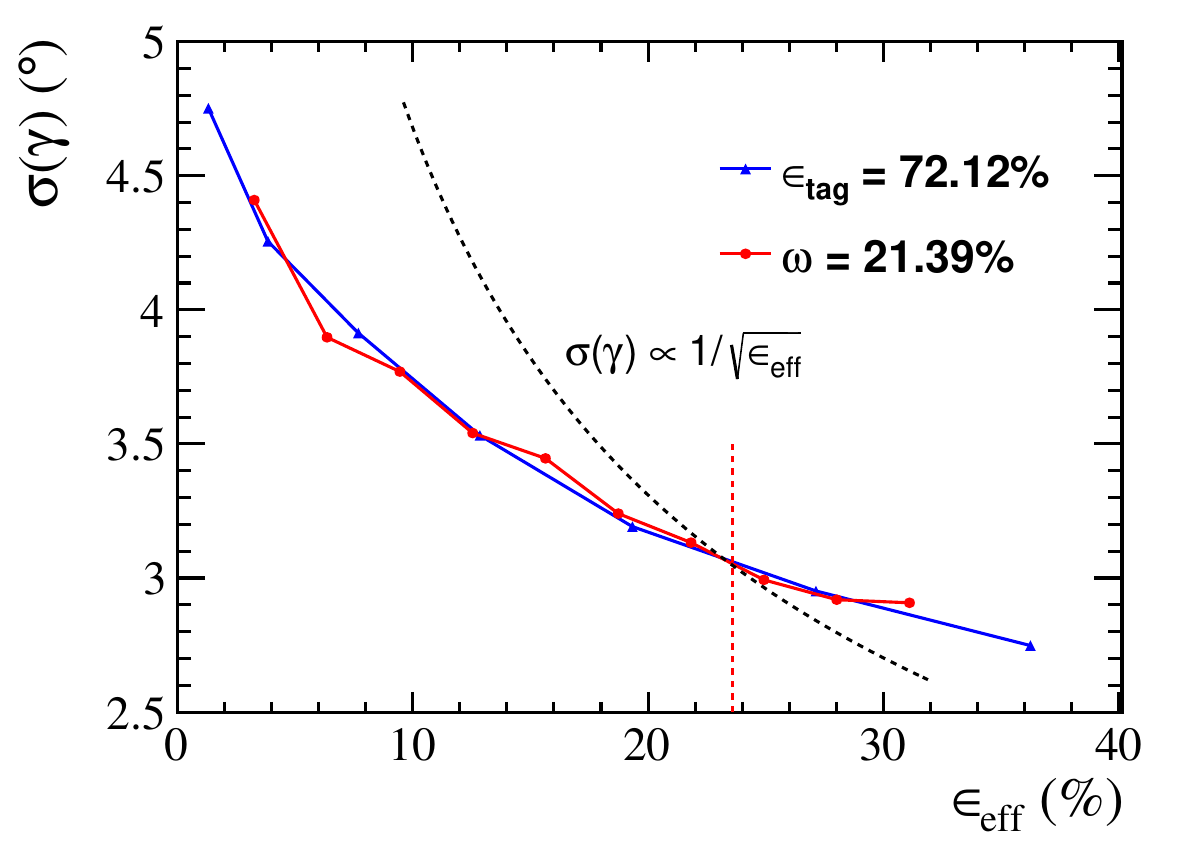}
\centering\includegraphics[width=0.45\textwidth]{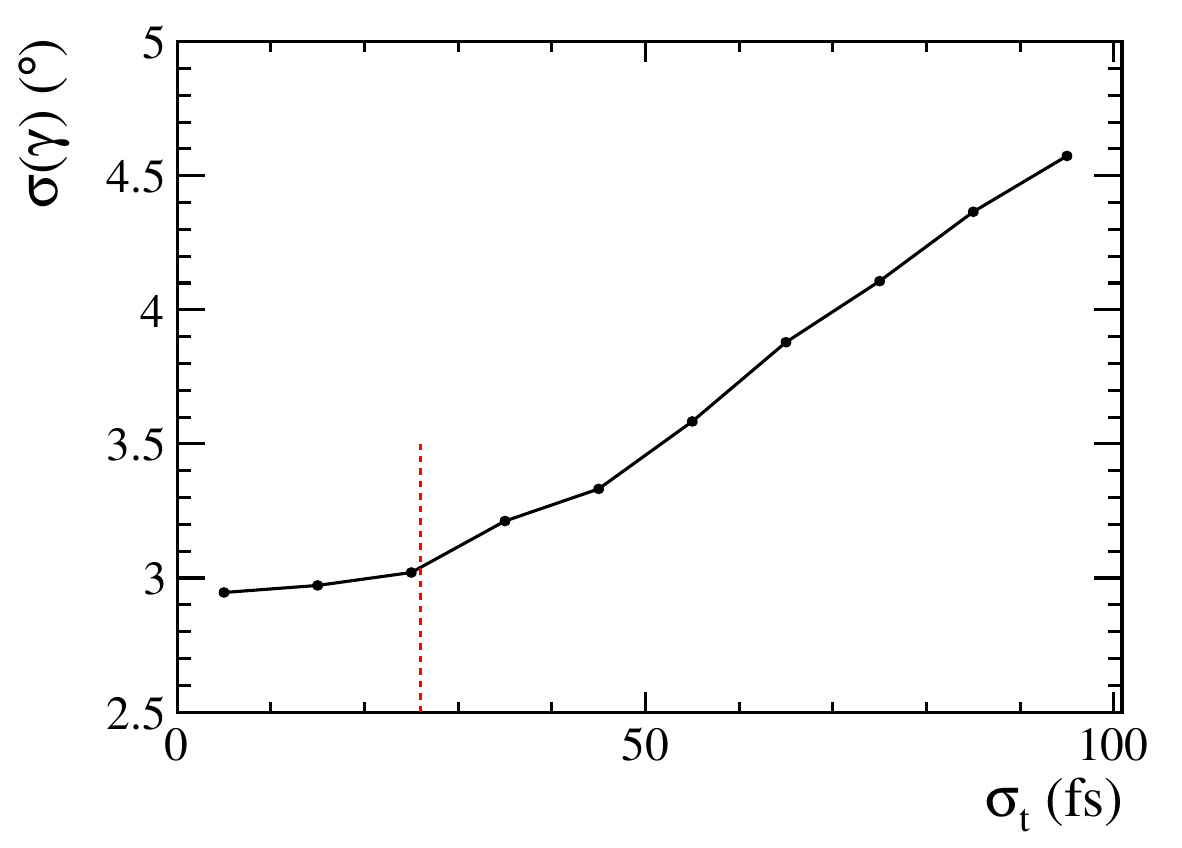}
\caption{Resolution of \(\gamma\) versus (left) flavor tagging power \(\epsilon_{\mathrm{eff}}\) and (right) effective decay time resolution \(\sigma_t\). Left: Dependence under fixed \(\epsilon_{\mathrm{tag}}\) (blue) vs fixed \(\omega\) (red), demonstrating \(\epsilon_{\mathrm{eff}}\) encapsulates the combined effects of these parameters, though deviations from the expected inverse square root dependence highlight the influence of untagged events. Right: Impact of \(\sigma_t\) variations, showing $26\mathrm{\,fs}$ resolution suffices. Vertical red dashed lines mark analysis baselines.}
\label{gammaErr_vs}
\end{figure}

The dependence of $\sigma(\gamma)$ on tagging efficiency and decay time resolution was investigated through toy Monte Carlo simulations. Fig.~\ref{gammaErr_vs} illustrates the evolution of $\gamma$ resolution as a function of two key parameters: tagging power $\epsilon_{\mathrm{eff}}$ and effective decay time resolution $\sigma_t$.

In the tagging efficiency analysis (left panel), two distinct operational scenarios were compared: The blue curve corresponds to a fixed tagging efficiency $\epsilon_{\mathrm{tag}} = 72.12\%$ with variable mistag fraction $\omega$, while the red curve represents a fixed $\omega = 21.39\%$ with varying $\epsilon_{\mathrm{tag}}$. The near-overlap of these curves demonstrates that tagging power $\epsilon_{\mathrm{eff}} \equiv \epsilon_{_\mathrm{tag}} (1-2\omega)^2$ successfully encapsulates the combined impact of tagging efficiency and mistag rate. However, the observed significant deviation from the expected inverse square root dependence ($\sigma(\gamma) \propto 1/\sqrt{\epsilon_{\mathrm{eff}}}$) reveals an important physical insight: $\gamma$ determination receives non-negligible contributions from untagged events, indicating that the conventional tagging power formulation does not fully account for all information channels in this specific analysis. 

For time resolution effects (right panel), we vary the dominant Gaussian component in the triple-Gaussian resolution model. The plot reveals that an effective time resolution of $26\mathrm{\,fs}$ provides sufficient precision for $\gamma$ extraction, with marginal gains observed for superior time resolution capabilities. This suggests the current analysis approach is near-optimal in terms of time resolution requirements.

\subsection{Full-statistics projection}
\label{sec:fullstats}

While current analysis focuses on partial statistics, extrapolating the $\gamma$ resolution to full statistics is crucial for evaluating its ultimate sensitivity. As mentioned in Sec.~\ref{sec:Chan}, the present analysis utilizes simulated events corresponding to 5.3\% of the combined expected yield for all three $D_s^-$ decay subchannels ($D_s^- \rightarrow K^-K^+\pi^-$, $K^-\pi^+\pi^-$, and $\pi^-\pi^+\pi^-$) under 4.1 Tera-$Z$ statistics. 

Following the well-established relationship $\sigma(\gamma) \propto 1/\sqrt{N_s}$, where the $\gamma$ resolution improves proportionally to the square root of the signal yield, we extrapolate the current subsample results to the full-statistics scenario. Assuming consistent experimental conditions, the projected $\gamma$ resolution under the CEPC 50~MW SR power operational scenario is determined to be $\sigma(\gamma) = {\gammaFinalErr}^{\circ}$.

\subsection{Optimization pathways}
Building upon the scaling framework established in~\cite{Li:2022tlo}, a series of optimization strategies are proposed corresponding to the parameters in the generalized scaling factor:
\begin{equation}
 \zeta = \left[N_{b\overline{b}} \cdot \mathrm{{\cal B}} \cdot \epsilon \cdot \left(\epsilon_{\mathrm{eff}} + \mathcal{D} \right) \cdot e^{-\mathcal{F} \cdot\Delta m^2_s \sigma^2_t }\right]^{-1/2}, 
\end{equation}
where $\mathcal{D}$ represents the deviation from the conventional tagging power and $\mathcal{F}$ denotes an empirical scaling factor.
First, as an electron-positron collider, CEPC can potentially reconstruct neutral pion with high precision, therefore, the inclusion of additional $D_s^-$ decay channels, especially those with neutral pions, such as $D_s^- \rightarrow K^- K^+ \pi^- \pi^0$ $({\cal B} = (5.50\pm0.24)$\%)\cite{PDG2024}, could significantly enhance the available statistics. 
Second, the current acceptance $\times$ efficiency product of $\epsilon = 80\%$ can be improved through the implementation of more sophisticated selection criteria. By incorporating machine learning techniques and multivariate analysis methods, the event selection process can be refined to achieve higher purity and efficiency. 
Third, while the present effective flavor tagging power of \( \epsilon_{\mathrm{eff}} = 23.6\% \) already surpasses typical hadronic collider performance, there exists considerable scope for improvement. Utilizing cutting-edge software tools, such as ParticleNet in tagging algorithms, could elevate this value to approximately \( 40\% \) \cite{Qu:2019gqs,Liang:2023wpt}.  
Fourth, the current effective decay time resolution of $\sigma_t = 26\mathrm{\,fs}$ presents another avenue for optimization. The implementation of advanced multivariate analysis techniques for vertex reconstruction could improve this parameter.

Furthermore, external constraints on the $B_s^0$ mixing parameters could be significantly tightened through dedicated $Z$ factory analyses. Examples include $\Delta m_s$ measurements via $B_s^0 \to D_s^- \pi^+$ decays, and $-2\beta_s$ determinations through the $B_s^0 \to J/\psi \phi$ channel. Recently, a study conducted at the Tera-$Z$ factory projected the resolution for the parameter \(-2\beta_s\) to be \(\sigma(-2\beta_s) = 4.3\)$\,\mathrm{mrad}$ \cite{Li:2022tlo}. 

The proposed optimization strategies, when implemented collectively, would lead to substantial improvements in the precision of the $\gamma$ measurement at CEPC.

\section{Conclusion}
We have performed a detailed investigation of $C\!P$-violating observables in $B_s^0 \to D_s^\mp K^\pm$ decays at a future Tera-$Z$ factory, demonstrating the experiment's exceptional potential for precision $\gamma$ measurements. The developed analysis framework incorporates realistic detector effects, including particle identification with $> 3\sigma$ $K/\pi$ separation, flavor tagging with 23.6\% effective efficiency, and decay time resolution modeling at $26\mathrm{\,fs}$. A two-dimensional simultaneous fit to mass and decay time distributions yields a statistical-only precision of $\sigma(\gamma) = \gammaAnaErr^\circ$ with partial statistics, projecting to $\sigma(\gamma) = \gammaFinalErr^\circ$ at full luminosity when combining all three dominant $D_s^-$ decay channels. 

This performance places CEPC at the forefront of $\gamma$-precision physics.  Future experiments aim to push boundaries: LHCb Upgrade I/II projects $\sigma(\gamma) = 2.5^\circ$/$1^\circ$ for this specific decay mode and $1^\circ$/$0.35^\circ$ in full combinations~\cite{LHCb:2018roe}. Strikingly, CEPC’s single-channel projection is already comparable to LHCb Upgrade II’s $B_s^0 \to D_s^\mp K^\pm$-specific goal and leverages only a fraction of its capabilities. By synergies with other golden modes (e.g., $B^\pm \to DK^\pm$), CEPC could surpass even the ultimate LHCb multi-channel precision of $0.35^\circ$. Such advances, enabled by the Tera-$Z$ factory’s clean collision environment and vertex resolution, would deliver definitive tests of CKM unitarity and uncover new physics through $C\!P$ violation anomalies. The demonstrated performance of the Tera-$Z$ factory strongly motivates its development as a 21st-century precision frontier machine.

\acknowledgments

We would like to thank Gang Li and Kaili Zhang for providing the inclusive $Z\rightarrow q\overline{q}$ samples, and are grateful to Hanhua Cui, Yongfeng Zhu and Zhijie Zhao for useful discussions.
We also thank Roy Aleksan and Emmanueal Francois Perez for the initial discussions.
This work is supported by the National Natural Science Foundation of China under grant No.12205312.

\providecommand{\href}[2]{#2}\begingroup\raggedright\endgroup

\bibliographystyle{JHEP}

\end{document}